\documentclass[11pt,a4paper]{article}

\usepackage{jheppub}

\newtheorem{lemma}{Lemma}[section]
\newtheorem{prop}{Proposition}[section]

\newcommand{\hook}
{\raisebox{-0.35ex}{\makebox[0.6em][r]{\scriptsize $-$}}
\hspace{-0.15em}\raisebox{0.25ex}{\makebox[0.4em][l]{\tiny $|$}}}

\newcommand{\be}{\begin{equation}}
\newcommand{\ee}{\end{equation}}
\newcommand{\ba}{\begin{eqnarray}}
\newcommand{\ea}{\end{eqnarray}}
\newcommand{\beq}{\begin{equation}}
\newcommand{\eeq}{\end{equation}}
\newcommand{\beqa}{\begin{eqnarray}}
\newcommand{\eeqa}{\end{eqnarray}}

\newcommand{\eq}[1]{(\ref{#1})}

\newcommand{\dt}{d^T\!}
\newcommand{\delt}{\delta^T\!}

\title{Local metrics admitting a principal Killing--Yano tensor with torsion}

\author[a]{Tsuyoshi Houri,}
\author[b]{David Kubiz\v{n}\'ak,}
\author[c]{Claude M.\ Warnick}
\author[d]{and Yukinori Yasui}

\affiliation[a]{%
Osaka City University Advanced Mathematical Institute (OCAMI),
3-3-138 Sugimoto, Sumiyoshi, Osaka, 558-8585, Japan}
\affiliation[b]{%
Perimeter Institute, 31 Caroline St.\ N.\ Waterloo Ontario, N2L 2Y5, Canada}
\affiliation[c]{%
Theoretical Physics Institute, University of Alberta, Edmonton, AB, T6G 2G7, Canada}
\affiliation[d]{%
Department of Mathematics and Physics, Graduate School of Science, Osaka City University,
3-3-138 Sugimoto, Sumiyoshi, Osaka, 558-8585, Japan}

\emailAdd{houri@sci.osaka-cu.ac.jp}
\emailAdd{dkubiznak@perimeterinstitute.ca}
\emailAdd{warnick@ualberta.ca}
\emailAdd{yasui@sci.osaka-cu.ac.jp}

\abstract{
In this paper we initiate a classification of local metrics admitting the principal 
Killing--Yano tensor with a skew-symmetric torsion.
It is demonstrated that in such spacetimes rank-2 Killing tensors occur naturally
and mutually commute. We reduce the classification problem to that of solving a set of partial differential equations, and we present some solutions to these PDEs.
In even dimensions, three types of local metrics are obtained:
one of them naturally generalizes the torsionless case
while the others occur only when the torsion is present.
In odd dimensions, we obtain more varieties of local
metrics.
The explicit metrics constructed in this paper are not the most general possible 
admitting the required symmetry, nevertheless, it is demonstrated that 
they cover a wide variety of solutions of various supergravities, such as
the Kerr-Sen black holes of (un-)gauged abelian heterotic supergravity, 
the Chong-Cvetic-L\"u-Pope black hole solution of five-dimensional minimal supergravity,
or the K\"ahler with torsion manifolds. The relation between generalized Killing--Yano tensors and various torsion Killing spinors 
is also discussed.}


\begin{document}
\hfill OCU-PHYS 361\\
\phantom{.} \hfill ALBERTA THY 4-12\\

\maketitle

\section{Introduction}
{Killing--Yano symmetry} has played an important role 
in the study of black hole physics
since Penrose and Floyd discovered that in the Kerr spacetime
a first integral of the geodesic equation can be written
as the square of a Killing--Yano tensor.
Killing--Yano tensors were first introduced from a purely mathematical point of view by Yano \cite{Yano:1952} 
and were later generalized to conformal Killing--Yano tensors 
by Tachibana and Kashiwada \cite{Tachibana:1969,Kashiwada:1968}. 
This symmetry is responsible for many remarkable properties 
of the Kerr geometry. Namely, it allows separation of variables for the Hamilton--Jacobi, Klein--Gordon, 
Dirac, and Maxwell equations in the curved Kerr background; solution of parallel transport equations; integration of stationary strings and
provides non-generic superinvariants for the supersymmetric spinning particle in this background.
Recently it was found that the existence of Killing--Yano symmetry extends to many higher-dimensional vacuua solutions of Einstein's equations with cosmological constant
describing rotating black holes with spherical horizon topology
\cite{Myers:1986,Gibbons:2004a,Chen:2006b}.
Due to the Killing--Yano symmetry, 
these higher-dimensional spacetimes possess 
similar integrability structures to the Kerr black hole, see, e.g., reviews \cite{Frolov:2008,Yasui:2011,Santillan:2011} and references therein.

Unfortunately, it turns out that the occurrence of standard Killing--Yano symmetry is rather limited---it is restricted to ``vacuum spacetimes'' of 
special algebraic type. This, for example, automatically disqualifies supergravity and string theory black hole solutions in the presence of 
fluxes. 
This lead the authors of \cite{Kubiznak:2009a, KubiznakEtal:2010} to introduce a notion of {\em generalized} conformal Killing--Yano symmetry 
where, in the simplest case, one extends the definition of Killing--Yano equations by considering the skew-symmetric torsion.
Such generalized symmetry naturally occurs 
in some higher-dimensional charged black hole spacetimes of supergravity theories,
while such spacetimes do not admit 
ordinary Killing--Yano symmetries \cite{Kubiznak:2009a, Houri:2010fr}.
For example, the black hole spacetime
of five-dimensional minimal gauged supergravity 
admits a Killing--Yano tensor with torsion, provided the torsion is identified with the Hodge dual of the Maxwell field:
${T}=*{F}/\sqrt{3}$.
This symmetry was also found in the Kerr--Sen black hole solution 
of effective string theory and its higher-dimensional generalizations, after identifying 
${T}$ with the 3-form field strength ${H}$.
In both cases, the generalized symmetry discovered in these spacetimes shares almost identical properties
with the standard Killing--Yano symmetry and implies the existence of important integrability structures for these black hole solutions.

Geometry with generalized Killing--Yano symmetry
is also related to the K\"ahler geometry 
studied by Apostolov, Calderbank and Gauduchon \cite{Apos:2006}.
In their paper, these authors introduced a notion of the Hamiltonian 2-form 
and obtained classification of all K\"ahler metrics with such a tensor.
These metrics can be also obtained as a BPS limit of Euclideanised higher-dimensional black hole spacetimes
\cite{Cvetic:2005a, Cvetic:2005b, Lu:2007, Chen:2006b, Chen:2007em, Hamamoto:2007, MartelliSparks:2005, Kubiznak:2009}.
We shall show that the generalized Killing--Yano symmetry arises on these K\"ahler manifolds and that it is responsible for separability of 
the Laplace operator therein. 

More generally, Killing--Yano symmetry appears naturally when one studies first-order symmetry operators of the Dirac operator with torsion \cite{HouriEtal:2010}; it emerges as a subset of necessary conditions for the existence of such an operator. Moreover, we shall show in this paper that, similar to the torsion-less case \cite{Semmelmann:2002}, various torsion Killing spinors give rise to a tower of all possible rank conformal Killing--Yano forms with torsion. Recently,
target spaces of supersymmetric non-relativistic particles with torsion 
were classified with the generalized Killing--Yano symmetry \cite{Papadopoulos:2011}---giving one more reason to study this generalized symmetry. 

In this paper, we attempt to classify spacetimes admitting
a {Killing--Yano} tensor with torsion. We derive explicit forms of the metrics and 
present some physically interesting examples.
When the torsion is absent, 
metrics with Killing--Yano symmetry
were classified in four dimensions \cite{Dietz:1981,Dietz:1982} 
and recently in higher dimensions \cite{Houri:2007,Houri:2008b,Houri:2009,Krtous:2008,
Papadopoulos:2008}.
It was shown in \cite{Houri:2008b,Krtous:2008} that a 
vacuum solution admitting the {\em principal Killing--Yano (PKY) tensor} (for definition see Sec. 2.2.) without torsion is uniquely given by the black hole metric found 
by Chen, L\"u and Pope \cite{Chen:2006b}.
Thus, it is a natural task to attempt to classify spacetimes admitting Killing--Yano symmetry 
when the torsion is present.
In particular, we concentrate on metrics admitting the {\em generalized PKY tensor}.
These metrics are expected to provide an {\em ansatz} 
for exact solutions of various supergravity theories \cite{CardosoEtAl:2012}. Hence our study provides an alternative to various approaches for finding new exact solutions, such as restricting to spacetimes with a sufficient number of isometries, supersymmetric spacetimes, spacetimes of special algebraic type, or spacetimes that can be written in a particular ansatz such as the Kerr--Schild form. 

Our strategy in classifying the metrics admitting the PKY tensor with torsion is to construct a canonical set of coordinates. Since the PKY tensor is a non-degenerate 2-form, it defines a canonical orthonormal frame at each point. Imposing that the 2-form satisfies the generalized PKY equation we are able to locally relate the canonical frame to a coordinate basis. In these coordinates many components of the torsion tensor vanish and we are left with a system of nonlinear partial differential equations whose solution gives rise to a metric admitting the PKY tensor with torsion. We are able to find large families of solutions of these equations in all dimensions, however, so far we have not been able to find an explicit general solution.

This paper is organized as follows:
In Sec.~2, we start with a brief review of conformal Killing--Yano tensors
with torsion, introduce the notion of generalized PKY tensor, and  
show that it generates the whole set of commuting rank-2 Killing tensors.
Sec.~3 and 4 represent the main body of the paper where we classify metrics admitting the PKY tensor with torsion. We demonstrate that 
there are three possible distinct types of metrics, 
which we call type A, B, and C.
Type A metrics can be regarded as a natural generalization of black hole spacetimes
and provide a unified description of several known solutions in supergravity theories.
Type B and C are exceptional metrics appearing only when the torsion is present.
In Sec.~5, we look for solutions of heterotic supergravity 
under the ansatz of type A metrics.
We find two types of solutions:
higher-dimensional Kerr-Sen black hole metrics and 
K\"ahler with torsion (KT) metrics including Calabi--Yau with torsion metrics. 
We believe that the latter are new.
We also show that in five dimensions the type A metric covers the Chong--Cvetic--L\"u--Pope black hole solution of five-dimensional minimal supergravity.
Sec.~6 is devoted to discussion and conclusions. In App.~A we discuss the relation of generalized Killing--Yano tensors to various torsion Killing spinors, App.~B collects
information about the Bismut connection, and App.~C gathers covariant derivatives of the canonical frame.

\section{Killing--Yano symmetry with torsion}
\subsection{Definition and basic properties}
We start with a review of Killing--Yano symmetry with torsion, see also \cite{Kubiznak:2009a, Houri:2010fr}.
Let $T$ be a 3-form on a $D$-dimensional Riemannian manifold $(M,g)$
and $\{e_a\}$ be an orthonormal frame, $g(e_a,e_b)=\delta_{ab}$.
We define a connection $\nabla^T$ by
\begin{align}
 \nabla^T_X Y
= \nabla_X Y + \frac{1}{2}\sum_a T(X,Y,e_a)\,e_a ~,
\end{align}
where $X$ and $Y$ are vector fields and $\nabla$ is the Levi-Civita connection. 
This connection satisfies a metricity condition, $\nabla^T g=0$, 
and has the same geodesics as $\nabla$, 
$\nabla^T_{\dot{\gamma}}\dot{\gamma}=\nabla_{\dot{\gamma}}\dot{\gamma}=0$ for a geodesic $\gamma$.
The connection 1-form ${\omega}^T{}^a{}_b$ is introduced by
\begin{align}
\nabla^T_{e_a}e_b = \sum_c {\omega}^T{}^c{}_b(e_a)\,e_c ~, \label{defcon}
\end{align}
which satisfies
\begin{align}
de^a+\sum_b {\omega}^T{}^a{}_b\wedge e^b = T^a \label{eq2-3}
\end{align}
where $T_a(X,Y)=T(e_a,X,Y)$.

For a $p$-form $\Psi$ a covariant derivative is calculated as
\begin{align}
\nabla^T_X\Psi 
= \nabla_X\Psi-\frac{1}{2}\sum_a(X \hook e_a\hook T)\wedge (e_a\hook \Psi) ~,
\end{align} 
where the operator $\hook$ represents the inner product.
Then, we have
\begin{align}
 d^T\Psi
=& \sum_a e^a \wedge\nabla^T_{e_a} \Psi \nonumber\\
=& d\Psi - \sum_a (e_a\hook T)\wedge(e_a\hook \Psi) ~, \\
 \delta^T \Psi 
=& -\sum_a e_a\hook \nabla^T_{e_a} \Psi \nonumber\\
=& \delta \Psi - \frac{1}{2}\sum_{a,b} 
   (e_a\hook e_b\hook T)\wedge(e_a \hook e_b\hook \Psi) ~.
\end{align}
For $\Psi=T$ one has $\delta^T T =\delta T$.

A {\it generalized conformal Killing--Yano (GCKY) tensor} $k$ introduced in \cite{Kubiznak:2009a} is 
a $p$-form satisfying for any vector field $X$ the following equation:
\begin{align}
\nabla^T_X k
= \frac{1}{p+1} X\hook d^T k 
  - \frac{1}{D-p+1} X^\flat\wedge\delta^T k ~, \label{1-5}
\end{align}
where $X^\flat$ is a dual 1-form of $X$.
We call a GCKY tensor $f$ obeying $\delta^T f=0$
a {\it generalized Killing--Yano tensor}, 
and a GCKY tensor $h$ obeying $d^T h=0$
a {\it $d^T\!$-closed GCKY tensor}.

We can see the GCKY equation as arising from representation theory considerations 
in the bundle of forms, cf. \cite{Semmelmann:2002,Moroianu:2003}. 
In general for a Riemannian manifold, 
one can decompose $T^*M \otimes \Lambda^p T^*M$ as an $O(n)$ representation as follows
\begin{equation}
T^*M \otimes \Lambda^p T^*M \cong \Lambda^{p+1}T^*M\oplus \Lambda^{p-1} T^*M \oplus \Lambda^{p,1} T^*M 
\end{equation}
where  $\Lambda^{p,1} T^*M$ consists of those elements $\alpha \otimes \psi$ 
of $T^*M \otimes \Lambda^p T^*M $ which satisfy $\alpha \wedge \psi = 0$, 
$\alpha^\sharp \hook \psi=0$. Applying this to $\nabla^T {k}$, 
one identifies the projection into $\Lambda^{p+1} T^*M$ as $d^T {k}$ 
and the projection into $\Lambda^{p-1} T^*M$ as $\delta^T {k}$, 
up to multiples. The generalized conformal Killing--Yano equation expresses 
the requirement that the component of $\nabla^T{k}$ transforming 
in the $\Lambda^{p,1}T^*M$ representation vanishes. 
The generalized Killing--Yano ($d^T\!$-closed GCKY) equation further requires that 
the component transforming in the $ \Lambda^{p-1} T^*M $ (resp.  $\Lambda^{p+1} T^*M $) vanishes. 
For this reason, we see that the existence of these tensors 
is closely tied to the underlying Riemannian geometry.

It is demonstrated in App. A that GCKY tensors arise naturally from corresponding torsion Killing spinors. 
For further general properties of these tensors we refer the reader to the paper \cite{Houri:2010fr}.

\subsection{Generalized PKY tensor} 
In what follows, we assume that $(M,g)$ 
admits a non-degenerate rank-2 $d^T\!$-closed GCKY tensor $h$ obeying
\begin{align}
 \nabla^T_X h = X^\flat\wedge \xi ~, \label{GCCKY Eq}
\end{align}
where
\begin{align}
 \xi=-\frac{1}{D-1}\delta^T h
\end{align}
is called an {\it associated 1-form of $h$}.
The terminology ``non-degenerate'' means that
the rank of $h$ as a (1,1)-tensor is maximal at all points of $M$
and that its eigenvalues are functionally independent.
We call a non-degenerate rank-2 $d^T\!$-closed GCKY tensor 
a {\it principal Killing--Yano (PKY) tensor with torsion}, 
or equivalently, a {\em generalized PKY tensor}.
Our aim is to classify spacetimes admitting this tensor.

In order to distinguish between even and odd dimensions we set $D=2n+\varepsilon$,
where $\varepsilon=0$ for even dimensions or $\varepsilon=1$ for odd dimensions,
and introduce a Darboux frame; 
$\{e^a\}=\{e^\mu,e^{\mu+n}=e^{\hat{\mu}}\}$ ($\mu=1,\cdots,n$) in even dimensions
and $\{e^a\}=\{e^\mu,e^{\mu+n}=e^{\hat{\mu}},e^{2n+1}=e^0\}$ in odd dimensions
in which the metric and the PKY tensor are written in the form 
(see, e.g., \cite{Frolov:2008} for the construction of this frame)
\begin{align}
g =& \sum_{\mu=1}^n (e^\mu\otimes e^\mu+e^{\hat{\mu}}\otimes e^{\hat{\mu}})
      +\varepsilon\,e^0\otimes e^0 ~, \label{formG} \\
h =& \sum_{\mu=1}^n x_\mu\,e^\mu\wedge e^{\hat{\mu}} ~. \label{formH}
\end{align}
Since there are still degrees of freedom under the rotation in each $(e^\mu,e^{\hat{\mu}})$-plane, 
we fix the orthonormal frame by taking the form of $\xi$ as, cf. \cite{Krtous:2008, Houri:2009},
\begin{align}
\xi = \sum_{\mu=1}^n\sqrt{Q_\mu}\,e^{\hat{\mu}}
           +\varepsilon\sqrt{Q_0}\,e^0 ~, \label{formxi}
\end{align}
where $Q_\mu$ and $Q_0$ are unknown functions.
This fully fixed orthonormal frame is called a {\it canonical frame}.
It follows from non-degeneracy of $h$ that $Q_\mu\neq 0$.
Although there is no reason why also the function $Q_0$ must be non-zero, 
we hereafter assume $Q_0\neq 0$.

Our task is to classify spacetimes with the generalized PKY tensor. 
This means that we have to determine not only all the possible metrics but also 
all the admissible torsions compatible with Eq.\ \eqref{GCCKY Eq} 
and the non-degeneracy of the PKY tensor. We will see 
in Sec.~3, that these requirements imply that many of 
the components of the torsion 3-form in the canonical frame must vanish.
More precisely, we obtain the following lemma:
\begin{lemma} \label{lemma21}
With respect to the canonical frame $\{e^a\}$, 
the torsion 3-form $T$ obeying (\ref{GCCKY Eq}) 
has only $\mu\hat{\mu}\hat{\nu}$-components ($\mu\neq\nu$)
in even dimensions while the other components are vanishing. 
In odd dimensions, the $\mu\hat{\mu}0$-components may also be non-zero.
That is, the torsion 3-form in $D=2n+\varepsilon$ dimensions takes the form
\begin{align}
 T = \sum_{\mu=1}^n\sum_{\nu\neq\mu}T_{\mu\hat{\mu}\hat{\nu}}
  \,e^\mu\wedge e^{\hat{\mu}}\wedge e^{\hat{\nu}}
  + \varepsilon \sum_{\mu=1}^n T_{\mu\hat{\mu}0}
  \,e^\mu\wedge e^{\hat{\mu}}\wedge e^0 ~. \label{formT}
\end{align}
\end{lemma}

As an immediate consequence of this lemma we infer the following:
Since $h$ is $d^T\!$-closed, we have
\begin{align}
 dh= \sum_a(e_a\hook T)\wedge(e_a\hook h) ~. \label{dh1}
\end{align}
Hence, by substituting (\ref{formH}) and (\ref{formT}) into (\ref{dh1}),
we obtain a relation between the PKY tensor and the torsion 3-form,
\begin{align}
 dh = -\sum_{\mu=1}^n \sum_{\nu\neq\mu}x_\mu T_{\nu\hat{\nu}\hat{\mu}}
   \,e^{\nu}\wedge e^{\hat{\nu}}\wedge e^\mu ~. \label{dh2}
\end{align}
This means that when in addition we require the PKY tensor to be closed, $dh=0$, 
then in even dimensions the torsion necessarily vanishes, while it can 
have only $\mu\hat{\mu}0$-components in odd dimensions.

Further geometrical interpretation is given in App.~B.

\subsection{Commuting Killing tensors}
It is known that in the absence of torsion, spacetimes admitting 
a PKY tensor have mutually commuting rank-2 Killing tensors
which are responsible for an integrable structure 
for the geodesic and Klein--Gordon equations.
We now show that the existence of such Killing tensors is also guaranteed when the torsion is present.

The following basic properties of GCKY tensors were demonstrated in \cite{Kubiznak:2009a, Houri:2010fr}:
\begin{enumerate}
\item A GCKY 1-form is equal to a conformal Killing 1-form.
In particular, a generalized Killing--Yano 1-form is equal to a Killing 1-form.
\item The Hodge star $*$ maps GCKY $p$-forms into GCKY ($D-p$)-forms.
In particular, the Hodge star of a $d^T\!$-closed GCKY $p$-form 
is a generalized Killing--Yano ($D-p$)-form and vice versa.
\item When $h_1$ and $h_2$ is a $d^T\!$-closed GCKY $p$-form and $q$-form, 
then $h_3=h_1\wedge h_2$ is a $d^T\!$-closed GCKY ($p+q$)-form.
\item Let $k$ be a generalized Killing--Yano $p$-form. Then the rank-2 symmetric tensor
\begin{align}
Q_{ab} = k_{ac_1\cdots c_{p-1}}k_b{}^{c_1\cdots c_{p-1}}
\end{align}
is a conformal Killing tensor.
In particular, $Q$ is a Killing tensor if $k$ is a generalized Killing--Yano tensor.
\end{enumerate}

By applying these properties to a PKY tensor ${h}$, 
the wedge products of $j$ PKY tensors, $h^{(j)}=h \wedge \cdots \wedge h$ ,
are rank-($2j$) $d^T\!$-closed GCKY tensors
and $f^{(j)}=* h^{(j)}$ are generalized Killing--Yano $(D-2j)$-forms.
In odd dimensions, $f^{(n)}=*h^{(n)}$ is a Killing vector.
Rank-2 symmetric tensors $K^{(j)}_{ab}=\left[j!^2(n-2j-1)!\right]^{-1}f^{(j)}_{ac_1\cdots c_{D-2j-1}} f_b^{(j)c_1\cdots c_{D-2j-1}}$ 
are Killing tensors, 
which are explicitly given by
\begin{align}
K^{(j)} = \sum_{\mu=1}^n A^{(j)}_{\mu}(e^\mu\otimes e^\mu
          +e^{\hat{\mu}}\otimes e^{\hat{\mu}}) \label{formK}
+\varepsilon\, A^{(j)}\,e^0\otimes e^0 ~,
\end{align}
where $A_\mu^{(k)}$ and $A^{(k)}$ are elementary symmetric polynomials in $x_\mu^2$ 
defined by the generating functions
\begin{align}
 \prod_{\nu=1}^n(t+x_\nu^2) 
=& A^{(0)}t^n+A^{(1)}t^{n-1}+\cdots +A^{(n)} ~, \\
 \prod_{\substack{\nu=1\\ \nu\neq\mu}}^n(t+x_\nu^2) 
=& A_\mu^{(0)}t^{n-1}+A_\mu^{(1)}t^{n-2}+\cdots +A_\mu^{(n-1)} ~. \label{defA}
\end{align}
We have the following proposition:

\begin{prop}
The Killing tensors $K^{(i)}$, \eqref{formK}, mutually commute
\begin{equation}
[ K^{(i)}, K^{(j)} ]=0 ~,
\end{equation}
under the Schouten--Nijenhuis bracket defined as 
\begin{align}
 [ K^{(i)}, K^{(j)} ]_{abc}
\equiv K^{(i)}_{e(a} \nabla^e K^{(j)}_{bc)}-K^{(j)}_{e(a} \nabla^e K^{(i)}_{bc)} ~.
\end{align}
\end{prop}
{\em Proof.} Using the connection $\nabla^T$, we rewrite $[ K^{(i)}, K^{(j)} ]_{abc}$ as
\begin{align}
K^{(i)}_{e(a} \nabla^e K^{(j)}_{bc)}-K^{(j)}_{e(a} \nabla^e K^{(i)}_{bc)}
=& K^{(i)}_{e(a} \nabla^{Te} K^{(j)}_{bc)}-K^{(j)}_{e(a} \nabla^{Te} K^{(i)}_{bc)} \nonumber\\
 & -K^{(i)}_{e(a}T^e{}_{b|d|}K^{(j)}{}^d{}_{c)}+K^{(j)}_{e(a}T^e{}_{b|d|}K^{(i)}{}^d{}_{c)} ~.
\end{align}
It was shown in \cite{Houri:2010fr} that these Killing tensors satisfy
$K^{(i)}_{e(a} \nabla^{Te} K^{(j)}_{bc)}-K^{(j)}_{e(a} \nabla^{Te} K^{(i)}_{bc)} =0$ and hence the first term on the r.h.s vanishes.
Using now the non-vanishing components of Killing tensors (\ref{formK}) and non-vanishing components of the torsion 3-form (\ref{formT}) derived in Lemma 2.1, 
it can be easily shown that also the second term on the r.h.s vanishes. $\Box$\\

To summarize, similar to the torsion-less case, the existence of the PKY tensor with torsion guarantees the existence of the whole tower of mutually commuting Killing tensors \eqref{formK}. However, contrary to the torsion-less case, it does not necessarily imply the existence of any Killing vector fields, except that in odd dimensions  
 $f^{(n)}=*h^{(n)}$ is a Killing vector.

\subsection{Integrability conditions}
Let us first note that from Eq. \eqref{formH} we have $x_\mu = {h}({e}_\mu,{e}_{\hat{\mu}})$.
By applying $\nabla^T_X$  to this relation we find
\begin{align}
  X(x_\mu) 
          =& \nabla^T_X h (e_\mu,e_{\hat{\mu}}) 
             + h (\nabla^T_X e_\mu, e_{\hat{\mu}}) 
             + h ({e}_\mu, \nabla^T_X e_{\hat{\mu}}) \nonumber\\
          =& g(X,e_\mu)\,g(\xi,e_{\hat{\mu}}) 
             - g(\xi,e_\mu)\,g(X,e_{\hat{\mu}}) ~,
\end{align}
which leads to
\begin{align}
 e_\nu(x_\mu) = \sqrt{Q_\mu}\,\delta_{\mu\nu} ~,~~~
 e_{\hat{\nu}}(x_\mu) = 0 ~,~~~
 e_0(x_\mu)=0 ~. \label{eq3-11}
\end{align}
The definition of  PKY (\ref{GCCKY Eq}) also implies 
\begin{align}
& \nabla^T_Y \nabla^T_X h 
  = (\nabla^T_Y X^\flat)\wedge \xi + X^\flat \wedge (\nabla^T_Y \xi) ~, \nonumber\\
& \nabla^T_{[X,Y]} h 
  = [X,Y]^\flat\wedge{\xi} ~.
\end{align}
Hence, we obtain the following integrability condition:
\begin{align}
\hat{R}^T({X},{Y}){h}
\equiv& (\nabla^T_X\nabla^T_Y-\nabla^T_Y\nabla^T_X-\nabla^T_{[X,Y]}){h} \nonumber\\
=& Y^\flat\wedge\nabla^T_X \xi -X^\flat\wedge\nabla^T_Y \xi 
  +\sum_c T(X,Y,e_c)\,e^c\wedge \xi  ~, \label{IC1}
\end{align}
where we have used $\nabla^T_X Y^\flat-\nabla^T_Y X^\flat-[X,Y]^\flat
=\sum_c T(X,Y,e_c)\,{e}^c$.
Since the curvature operator $\hat{R}^T({X},{Y})$ is related to 
the curvature with torsion ${\cal R}^T$ 
by ${\cal R}^T(X,Y,Z,W)
=g(\hat{R}^T(X,Y)Z,W)$, in components the integrability condition reads 
\begin{align}
 {\cal R}^T_{abc}{}^eh_{ed}-{\cal R}^T_{abd}{}^eh_{ec}
 =& g_{ad}\nabla^T_b\xi_c -g_{ac}\nabla^T_b\xi_d -g_{bd}\nabla^T_a\xi_c + g_{bc}\nabla^T_a\xi_d\nonumber\\
  &+T_{abc}\xi_d-T_{abd}\xi_c ~. \label{IC2}
\end{align}
Both equations \eqref{eq3-11} and \eqref{IC2} will be exploited in the next section.

\section{Classification of metrics admitting a PKY tensor}
Let us first discuss possible local forms of metrics in {\em even dimensions}; 
classification of odd-dimensional metrics is deferred to Sec.~4.
We proceed as follows: By applying the techniques developed in \cite{Krtous:2008, Houri:2009}, we first 
restrict the form of the connection 1-forms with torsion. Employing further the PKY equation \eqref{GCCKY Eq} and its 
integrability conditions we can eliminate a great number of unknown components of the torsion. In addition, we can determine
covariant derivatives $\nabla^T_{e_a}{e}_b$ in terms of the eigenvalues $x_\mu$, the unknown functions $Q_\mu$, and derivatives of the associated 1-form 
${\xi}$. Finally, we study commutators of the basis vectors. Imposing the Jacobi identity, we are able to further restrict the form of the canonical frame, we derive necessary differential constraints \eqref{K1}---\eqref{originaleven}, and establish an important ``algebraic relation'', \eqref{MK}, which we use for the classification of all possible metrics. We divide such metrics into type A, B, and C, and study them in the appropriate subsections.

\subsection{Connection 1-forms}
To restrict the form of connection 1-forms with torsion we employ the techniques developed in \cite{Krtous:2008, Houri:2009}. Namely, we define 
a ($1,1$)-tensor ${Q}$ as 
\begin{align}
 Q^a{}_b = -h^a{}_ch^c{}_b ~,
\end{align}
and denote its spectrum by $\lambda_i=\{x_1^2, x_2^2,\dots, x_n^2, 0\}$, where 
the eigenvalues $x_\mu^2$ are of multiplicity two, and the last zero value eigenvalue $\lambda_{n+1}=0$ is present only in odd dimensions. Hence we take,  
$i=1,\cdots,n+\varepsilon$. 
We also introduce the orthogonal projection operators ${\cal P}(\lambda_i)$ 
which map a vector onto its component in the eigenspace of $\lambda_i$. In particular, ${Q}=\sum_i \lambda_i P(\lambda_i)$, and, from here,
\begin{align}
({I}+t{Q})^{-1}=\sum_{i=1}^{n+\varepsilon}\frac{1}{1+t\lambda_i}{\cal P}(\lambda_i) ~.
\label{projection}
\end{align}
By differentiating both sides of Eq.\ (\ref{projection}) 
and using the PKY equation,
the covariant derivatives of the projection operators
can be evaluated as follows \cite{Houri:2009}:
\be
\nabla^T_a{\cal P}(x_\mu^2)_{bc}
=  \sum_{\nu\neq\mu}
   \frac{F(x_\mu^2,x_\nu^2)_{abc}}{x_\mu^2-x_\nu^2}
   +\frac{F(x_\mu^2,0)_{abc}}{x_\mu^2} ~,\quad
\nabla^T_a{\cal P}(0)_{bc}
=  -\sum_{\mu=1}^n\frac{F(0,x_\mu^2)_{abc}}{x_\mu^2} ~,
\ee
where 
\ba
F(x_\mu^2,x_\nu^2)_{abc}
&=& x_\nu\sqrt{Q_\nu}{\cal P}(x_\mu^2)_{ab}(e^\nu)_c
  +\sqrt{Q_\nu}h_{ad}{\cal P}(x_\mu^2)^d{}_b(e^{\hat{\nu}})_c \nonumber\\
& & +x_\mu\sqrt{Q_\mu}{\cal P}(x_\nu^2)_{ab}(e^\mu)_c
  +\sqrt{Q_\mu}h_{ad}{\cal P}(x_\nu^2)^d{}_b(e^{\hat{\mu}})_c
  ~+~(b \leftrightarrow c) ~, \label{qa1} \\
F(x_\mu^2,0)_{abc}
&=& \sqrt{Q_0}h_{ad}{\cal P}(x_\mu^2)^d{}_b(e^0)_c +x_\mu\sqrt{Q_\mu}{\cal P}(0)_{ab}(e^\mu)_c
   ~+~ (b \leftrightarrow c) ~. \label{qa2}\nonumber
\ea
From (\ref{formH}) and (\ref{formxi}), the canonical bases are written as
\be
 ({e}_\mu)^a = \frac{1}{x_\mu\sqrt{Q_\mu}}h^a{}_b{\cal P}(x_\mu^2)^b{}_c\xi^c ~,\quad
 ({e}_{\hat{\mu}})^a = \frac{1}{\sqrt{Q_\mu}}{\cal P}(x_\mu^2)^a{}_b\xi^b ~, \quad
 ({e}_0)^a = \frac{1}{\sqrt{Q_0}}{\cal P}(0)^a{}_b\xi^b ~. \label{orthopro}
\ee
Using the equation $ {\omega}^a{}_b({e}_c) = {e}^a(\nabla^T_c{e}_b)$, 
let us calculate $({\omega}^T)^\mu{}_{\hat{\mu}}$ as follows:
\ba
{\omega}^T{}^\mu{}_{\hat{\mu}}({e}_a)
&=& ({e}^\mu)_b\nabla^T_a({e}_{\hat{\mu}})^b =({e}^\mu)_b\nabla^T_a
   \Big(\frac{1}{\sqrt{Q_\mu}}{\cal P}(x_\mu^2)^b{}_c\xi^c\Big)\nonumber\\
&=& ({e}^\mu)_b\Big(\nabla^T_a\frac{1}{\sqrt{Q_\mu}}\Big){\cal P}(x_\mu^2)^b{}_c\xi^c
   +\frac{1}{\sqrt{Q_\mu}}({e}^\mu)_b\Big(\nabla^T_a{\cal P}(x_\mu^2)^b{}_c\Big)\xi^c 
 \nonumber\\
& & +\frac{1}{\sqrt{Q_\mu}}({e}^\mu)_b{\cal P}(x_\mu^2)^b{}_c\Big(\nabla^T_a\xi^c\Big) \nonumber\\
&=& \sum_{\nu\neq\mu}
    \frac{Q_\nu h_{ab}{\cal P}(x_\mu^2)^b{}_c({e}_\mu)^c}{\sqrt{Q_\mu}(x_\mu^2-x_\nu^2)}
   +\sum_{\nu\neq\mu}
    \frac{\sqrt{Q_\mu}h_{ab}{\cal P}(x_\nu^2)^b{}_c({e}_\mu)^c}{x_\mu^2-x_\nu^2}
   +\frac{Q_0h_{ab}{\cal P}(x_\mu^2)^b{}_c({e}_\mu)^c}{x_\mu^2\sqrt{Q_\mu}} \nonumber\\
& &+\sum_{\nu\neq\mu}\frac{x_\mu\sqrt{Q_\nu}}{x_\mu^2-x_\nu^2}
    ({e}^{\hat{\nu}})_a 
   +\varepsilon\,k\frac{\sqrt{Q_0}}{x_\mu}({e}^0)_a
   +\frac{1}{\sqrt{Q_\mu}} ({e}^\mu)_c\nabla^T_a\xi^c ~.
\ea
Thus, the following connection 1-forms are obtained:
\begin{align}
{\omega}^T{}^\mu{}_{\hat{\mu}}
=& \frac{1}{\sqrt{Q_\mu}}\Bigg(
   -\sum_{\substack{\nu=1 \\ \nu\neq\mu}}^n\frac{x_\mu Q_\nu}{x_\mu^2-x_\nu^2}-\frac{Q_0}{x_\mu} 
   \Bigg)\,{e}^{\hat{\mu}} 
   +\sum_{\substack{\nu=1 \\ \nu\neq\mu}}^n\frac{x_\mu\sqrt{Q_\nu}}{x_\mu^2-x_\nu^2}\,{e}^{\hat{\nu}} \nonumber\\
 &   +\varepsilon\,\frac{\sqrt{Q_0}}{x_\mu}\,{e}^0 
   +\frac{1}{\sqrt{Q_\mu}}\sum_a\Big(({e}^\mu)_c\nabla^T_a\xi^c\Big)\,{e}^a ~.
\end{align}
The other connection 1-forms are calculated similarly. In particular, in even dimensions we obtain the following: 
\begin{lemma}
In even dimensions the connection 1-forms with torsion must have the following form:
\begin{align}\label{omega1}
{\omega}^T{}^\mu{}_\nu
=& -\frac{x_\nu\sqrt{Q_\nu}}{x_\mu^2-x_\nu^2}\,{e}^\mu
   -\frac{x_\mu\sqrt{Q_\mu}}{x_\mu^2-x_\nu^2}\,{e}^\nu ~, ~~~~~(\mu\neq\nu) \nonumber\\
{\omega}^T{}^\mu{}_{\hat{\mu}}
=& -\frac{1}{\sqrt{Q_\mu}}
   \sum_{\nu\neq\mu}\frac{x_\mu Q_\nu}{x_\mu^2-x_\nu^2}
   \,{e}^{\hat{\mu}} 
   +\sum_{\nu\neq\mu}
   \frac{x_\mu\sqrt{Q_\nu}}{x_\mu^2-x_\nu^2}\,{e}^{\hat{\nu}}
   +\sum_a\frac{\kappa_a{}^\mu}{\sqrt{Q_\mu}}\,{e}^a ~, \nonumber 
\\
{\omega}^T{}^\mu{}_{\hat{\nu}}
=& \frac{x_\mu\sqrt{Q_\nu}}{x_\mu^2-x_\nu^2}\,{e}^{\hat{\mu}} 
    -\frac{x_\mu\sqrt{Q_\mu}}{x_\mu^2-x_\nu^2}\,{e}^{\hat{\nu}} ~, ~~~~~(\mu\neq\nu) 
\\
{\omega}^T{}^{\hat{\mu}}{}_{\hat{\nu}}
=& -\frac{x_\mu\sqrt{Q_\nu}}{x_\mu^2-x_\nu^2}\,{e}^\mu
    -\frac{x_\nu\sqrt{Q_\mu}}{x_\mu^2-x_\nu^2}\,{e}^\nu ~, ~~~~~(\mu\neq\nu) \nonumber 
\end{align}
where 
\begin{align}
\kappa_a{}^b \equiv ({e}^b)_c\nabla^T_a{\xi}^c ~. \label{kappadef}
\end{align}
\end{lemma}
The components of connection 1-forms are expressed in terms of the  
 eigenvalues $x_\mu$ of the PKY tensor, the components $\sqrt{Q_\mu}$ of the associated 1-form, and $\kappa_{ab}$. 
 
We can gain more information about $\kappa_{ab}$ by directly differentiating (\ref{formxi}) and using (\ref{omega1}), to 
obtain
\begin{align}
& \kappa_{\mu\hat{\mu}}=
  {e}_\mu(\sqrt{Q_\mu})-\sum_{\nu\neq\mu}\frac{x_\mu Q_\nu}{x_\mu^2-x_\nu^2} ~, \label{kappa1} \\
& \kappa_{\mu\hat{\nu}}=
  {e}_\mu(\sqrt{Q_\nu})+\frac{x_\mu\sqrt{Q_\nu}\sqrt{Q_\mu}}{x_\mu^2-x_\nu^2} ~,
  ~~~~~(\mu\neq\nu) \label{kappa2} \\
& \kappa_{\hat{\mu}\hat{\mu}}={e}_{\hat{\mu}}(\sqrt{Q_\mu}) ~, \label{kappa3} \\[0.2cm]
& \kappa_{\hat{\mu}\hat{\nu}}={e}_{\hat{\mu}}(\sqrt{Q_\nu}) ~.
\qquad (\mu\neq\nu) \label{kappa4}
\end{align}
Also, by evaluating the integrability condition 
(\ref{IC2}) on $(c,d)=(\mu,\hat{\mu})$ and
using the fact that ${h}$ is diagonalized in the canonical frame, it follows that
\begin{align}
\delta_{a\hat{\mu}}\kappa_{b\mu}-\delta_{a\mu}\kappa_{b\hat{\mu}}
-\delta_{b\hat{\mu}}\kappa_{a\mu}+\delta_{b\mu}\kappa_{a\hat{\mu}}
+\sqrt{Q_\mu}T_{ab\mu} = 0 ~.
\end{align}
Hence we find
\begin{align}
& \kappa_{\mu\mu} +\kappa_{\hat{\mu}\hat{\mu}}=0 ~, \label{eq310}\\
& \kappa_{\mu\nu} =\kappa_{\mu\hat{\nu}}
  =\kappa_{\hat{\mu}\hat{\nu}}=0 ~,
  ~~~~~(\mu\neq\nu) \label{eq311}\\
& \kappa_{\hat{\mu}\nu}=-\sqrt{Q_\nu}T_{\hat{\mu}\nu\hat{\nu}} ~,
  ~~~~~(\mu\neq\nu) \label{eq312}
\end{align}
and obtain
\begin{align}
& T_{\mu\nu\hat{\nu}}= 0 ~,
  ~~~~~(\mu\neq\nu) \label{eq313-1} \\
& T_{\mu\nu\rho}=T_{\mu\nu\hat{\rho}}=T_{\mu\hat{\nu}\hat{\rho}}= 0 ~.
  ~~~~~(\text{$\mu,\nu,\rho$ all different}) \label{eq313-2}
\end{align}

From the connection 1-forms (\ref{omega1}) together with (\ref{eq311})
one can evaluate covariant derivatives $\nabla^T_{e_a}{e}_b$, 
which are summarized in App.~C.
Using these expressions and Eq. \eqref{eq3-11}, we can directly confirm that 
(\ref{formH}) satisfies the PKY equation (\ref{GCCKY Eq}).

\subsection{Commutators}
To obtain yet more information about $\kappa^a{}_b$, 
we consider the commutation relations.
Using 
\begin{align}
 [{e}_a,{e}_b] = \nabla^T_{e_a}{e}_b-\nabla^T_{e_b}{e}_a 
                       - \sum_c {T}({e}_a,{e}_b,{e}_c){e}_c ~, \label{3.16}
\end{align}
we have
\ba
\left[ {e}_\mu, {e}_\nu \right] 
&=& -\frac{x_\nu\sqrt{Q_\nu}}{x_\mu^2-x_\nu^2}\,{e}_\mu
    -\frac{x_\mu\sqrt{Q_\mu}}{x_\mu^2-x_\nu^2}\,{e}_\nu ~, ~~~~~(\mu\neq\nu) \label{comev1}\\
\left[ {e}_\mu, {e}_{\hat{\mu}} \right]
&=& K_\mu\,{e}_\mu
    +L_\mu\,{e}_{\hat{\mu}} 
    +\sum_{\rho\neq\mu} M_{\mu\rho} \,{e}_{\hat{\rho}} ~,\\
\left[{e}_\mu,{e}_{\hat{\nu}}\right]
&=&-\frac{x_\mu\sqrt{Q_\mu}}{x_\mu^2-x_\nu^2}\,{e}_{\hat{\nu}} ~,~~~~~ (\mu\neq\nu) \\
\left[{e}_{\hat{\mu}},{e}_{\hat{\nu}}\right]
&=& -\sum_{\rho\neq\mu,\nu}T_{\hat{\mu}\hat{\nu}\hat{\rho}}\,{e}_{\hat{\rho}} ~,~~~~~ (\mu\neq\nu) \label{comev4}
\ea
where we have defined
\ba
K_\mu &\equiv& \frac{\kappa_\mu{}^\mu}{\sqrt{Q_\mu}}=- \frac{{e}_{\hat{\mu}}(\sqrt{Q_\mu})}{\sqrt{Q_\mu}}\,,\label{Kmu}\\
L_\mu &\equiv& -\frac{1}{\sqrt{Q_\mu}}\Bigg(\sum_{\rho\neq\mu}
  \frac{x_\mu Q_\rho}{x_\mu^2-x_\rho^2}-\kappa_{\hat{\mu}}{}^\mu\Bigg) ~, \\
M_{\mu\nu} &\equiv& \frac{2x_\mu\sqrt{Q_\nu}}{x_\mu^2-x_\nu^2}
  -T_{\mu\hat{\mu}\hat{\nu}} ~, ~~~~~(\mu\neq\nu) \label{KLMdef}
\ea
and we have used (\ref{kappa3}) and (\ref{eq310}).

We can demonstrate that a new frame $\{{\epsilon}_\mu\}$ 
defined by ${\epsilon}_\mu={e}_\mu/\sqrt{Q_\mu}$ satisfies 
$[{\epsilon}_\mu,{\epsilon}_\nu]=0$.
From Frobenius' theorem, therefore, we can choose $x_\mu$ as local coordinates
of an integral submanifold ${\cal N}$ and 
the vector fields ${e}_\mu$ can be locally written as (see also \cite{Krtous:2008})
\begin{align}
{e}_\mu = \sqrt{Q_\mu}{\frac{\partial}{\partial x_\mu}} ~. \label{2-108}
\end{align}
Furthermore, together with (\ref{kappa2}), (\ref{kappa4}) and (\ref{eq311}), 
we can determine the form of the functions $Q_\mu$ as follows:
\begin{align}
Q_\mu = \frac{X_\mu}{U_\mu} ~,~~~~~
U_\mu = \prod_{\nu\neq\mu} (x_\mu^2-x_\nu^2) ~, \label{Qmu}
\end{align}
where $X_\mu$ are arbitrary functions satisfying ${e}_\nu(X_\mu)={e}_{\hat{\nu}}(X_\mu)=0$
for $\nu\neq\mu$.

We have restricted the forms of the connection 1-forms 
by essentially using the integrability condition of the PKY tensor.
However, Eqs.\ (\ref{comev1})--(\ref{comev4}) do not yet satisfy the Jacobi identity
\begin{align}
[[{e}_a,{e}_b],{e}_c]
+[[{e}_b,{e}_c],{e}_a]
+[[{e}_c,{e}_a],{e}_b]=0 ~,
\end{align}
which is equivalent to the first Bianchi identity.
After some calculations, we find that
the $\hat{\mu}\hat{\nu}\hat{\rho}$-components ($\mu,\nu,\rho$ all different) 
of the torsion 3-form must vanish, $T_{\hat{\mu}\hat{\nu}\hat{\rho}}=0$.
Thus, combining this result with (\ref{eq313-1}) and (\ref{eq313-2}), we obtain Lemma~\ref{lemma21}.
Now we have $[{e}_{\hat{\mu}},{e}_{\hat{\nu}}]=0$,
which provides an integrable distribution spanned by ${e}_{\hat{\mu}}$
aside from the previous integrable distribution $\cal N$.
Simultaneously, the Jacobi identities require the algebraic equation
\begin{align}
 M_{\mu\nu}K_\nu = 0 ~,~~~~~(\mu\neq\nu,~\mbox{no sum)} \label{MK}
\end{align}
and the following system
of partial differential equations for $K_\mu$, $L_\mu$ and $M_{\mu\nu}$ 
($\mu,\nu,\rho$ all different and no sum):
\begin{align}
& \partial_\nu K_\mu = \frac{x_\nu K_\mu}{x_\mu^2-x_\nu^2} ~, \label{K1} \\
& \partial_\nu L_\mu = \frac{x_\nu L_\mu}{x_\mu^2-x_\nu^2}
                        -\frac{M_{\mu\nu}M_{\nu\mu}}{\sqrt{Q_\nu}}
  -\frac{2x_\mu x_\nu\sqrt{Q_\mu}}{(x_\mu^2-x_\nu^2)^2} ~, \label{L1}\\
& \partial_\nu M_{\mu\nu} = \Bigg(\frac{2x_\nu}{x_\mu^2-x_\nu^2}-\frac{L_\nu}{\sqrt{Q_\nu}}\Bigg)M_{\mu\nu} ~, 
 \label{M1}\\
& \partial_\nu M_{\mu\rho} = \Bigg(\frac{2x_\nu}{x_\mu^2-x_\nu^2}
                              +\frac{x_\nu}{x_\nu^2-x_\rho^2}\Bigg)M_{\mu\rho} 
                              -\frac{M_{\mu\nu}M_{\nu\rho}}{\sqrt{Q_\nu}} ~, \label{M2}
\end{align}
and
\begin{align}
  {e}_{\hat{\nu}}(K_\mu) = 0 ~,~~~ 
  {e}_{\hat{\nu}}(L_\mu) = 0 ~,~~~
  {e}_{\hat{\nu}}(M_{\mu\nu}) = 0 ~,~~~
  {e}_{\hat{\nu}}(M_{\mu\rho}) = 0 ~. \label{originaleven}
\end{align}
From Eq.\ (\ref{MK}), one finds that there are three types of solutions:
{(type A)} $K_\mu=0$ for all $\mu$, {(type B)} $M_{\mu\nu}=0$ for all $\mu,\nu$, 
and {(type C) Mixed case}, i.e., $K_\mu\neq0$ for $\mu=1,\cdots,k$ ($1<k<n$) 
and $K_\mu=0$ for $\mu=k+1,\cdots,n$.
Eq.\ (\ref{K1}) automatically holds for (\ref{Kmu}).
The integrability conditions of (\ref{L1}), (\ref{M1}) and (\ref{M2}) are satisfied,
namely differentiating these equations does not produce any additional equations.
Note that Eqs.\ (\ref{comev1})--(\ref{comev4}) are easily written as
\begin{align}
d{e}^\mu
=& \sum_{\nu\neq\mu}\frac{x_\nu\sqrt{Q_\nu}}{x_\mu^2-x_\nu^2}\,{e}^\mu\wedge{e}^\nu
   -K_\mu\,{e}^\mu\wedge{e}^{\hat{\mu}} ~,  \label{3.32} \\
d{e}^{\hat{\mu}}
=& -L_\mu\,{e}^\mu\wedge{e}^{\hat{\mu}} 
   -\sum_{\nu\neq\mu}\frac{x_\nu\sqrt{Q_\nu}}{x_\mu^2-x_\nu^2}\,{e}^\nu\wedge{e}^{\hat{\mu}}
   -\sum_{\nu\neq\mu}M_{\nu\mu}\,{e}^\nu\wedge{e}^{\hat{\nu}} ~. \label{3.33}
\end{align}
Thus our problem has been reduced to finding the solutions to 
(\ref{L1})--(\ref{originaleven}), 
and then finding the canonical frame $\{{e}^a\}$ 
obeying (\ref{3.32}) and (\ref{3.33}).

\subsection{Type A: $K_\mu=0$ case}
Let us first consider the case of $K_\mu=0$ for all $\mu$.
For simplicity, we assume that functions $L_\mu$ and $M_{\mu\nu}$ 
depend only on $x_\mu$-coordinate, so that Eqs.\ (\ref{originaleven}) are trivially satisfied
since ${e}_{\hat{\mu}}(x_\nu)=0$ for all $\mu,\nu$.
Eq.\ (\ref{Kmu}) shows that ${e}_{\hat{\mu}}(\sqrt{Q_\mu})=0$ for all $\mu$,
which, together with  ${e}_{\hat{\nu}}(\sqrt{Q_\mu})=0$, implies that  
functions $X_\mu$ are functions of one variable only:
$X_\mu = X_\mu(x_\mu)$.

For Eqs. (\ref{L1}), (\ref{M1}) and (\ref{M2}),
we obtain the following solution:\footnote{
In the absence of torsion, the general solution is 
$L_\mu=-\partial_\mu\sqrt{Q_\mu}$ and 
$M_{\mu\nu} = 2x_\mu\sqrt{Q_\nu}/(x_\mu^2-x_\nu^2)$, 
which leads to the Kerr-NUT-(A)dS spacetimes found in \cite{Chen:2006b}.}
\begin{align}
& L_\mu = -\partial_\mu\sqrt{Q_\mu}
         +\Big(\partial_\mu \ln \frac{\Phi}{f_\mu}\Big)\sqrt{Q_\mu} ~, \label{sol31} \\
& M_{\mu\nu} = \frac{f_\nu}{f_\mu}\Big(\frac{2x_\mu}{x_\mu^2-x_\nu^2}
               +\partial_\mu \ln \Phi\Big)\sqrt{Q_\nu} ~, \label{sol32}
\end{align}
where $\Phi$ is a function obeying $\partial_\mu\partial_\nu \Big[(x_\mu^2-x_\nu^2)\Phi\Big] = 0\,$, and can 
be solved in the form\footnote{
Here, 1 is just a convenient choice of normalization for the integration constant.
Choosing different value would slightly change the final expression (\ref{A1even}).}
\begin{align}
 \Phi = 1 + \sum_{\mu=1}^n\frac{N_\mu}{U_\mu} ~. \label{defH}
\end{align}
Thus our solution includes $3n$ arbitrary functions $X_\mu$, $f_\mu$ and $N_\mu$ 
depending on one variable $x_\mu$ only.
From Lemma 2.1, (\ref{KLMdef}) and (\ref{sol32}),
the torsion 3-form is given by
\begin{align}
{T}
  = \sum_{\mu\neq\nu}\Bigg[\frac{2x_\mu}{x_\mu^2-x_\nu^2}
    -\frac{f_\nu}{f_\mu}\Big(\frac{2x_\mu}{x_\mu^2-x_\nu^2}
    +\partial_\mu\ln \Phi\Big)\Bigg]\sqrt{Q_\nu}
    \,{e}^\mu\wedge {e}^{\hat{\mu}}\wedge{e}^{\hat{\nu}} ~. \label{tsnevenA}
\end{align}

Finally we have to solve Eqs. (\ref{3.32}) and (\ref{3.33}).
This is done as follows.
It is possible to show that the the following 2-form:
\begin{align}
{F}_{(2)} = \sum_{\mu=1}^n\frac{\partial_\mu\ln \Phi}{f_\mu}\,{e}^\mu\wedge{e}^{\hat{\mu}} \label{2-form}
\end{align}
is $d$-closed and hence can be locally written as ${F}_{(2)}={d}{A}_{(1)}$. 
Furthermore, we can prove that 1-forms
\begin{align}
{\theta}_k
  = \sum_{\mu=1}^n\frac{(-1)^kx_\mu^{2(n-k-1)}}{U_\mu}
    \frac{{e}^{\hat{\mu}}}{f_\mu\sqrt{Q_\mu}}
    +\delta_{k0}\,{A}_{(1)} ~,~~~~~
  k=0,\cdots,n-1 \label{thetak}
\end{align}
are also $d$-closed. 
We can introduce local functions $\psi_k$ such that ${\theta}_k={d}\psi_k$.
Thus the canonical frame reads
\begin{align}
{e}^{\mu} = \frac{dx_\mu}{\sqrt{Q_\mu}} ~,~~~
{e}^{\hat{\mu}} = f_\mu\sqrt{Q_\mu}
\Bigg(\sum_{k=0}^{n-1}A_\mu^{(k)}d\psi_k-{A}_{(1)}\Bigg) ~, \label{LF1}
\end{align}
where 1-form ${A}_{(1)}$ takes the form
${A}_{(1)}=d\psi_0 + \sum_{\mu=1}^n A_{\hat{\mu}}\,{e}^{\hat{\mu}}$.
By exploiting the gauge freedom we can eliminate the exact term, to obtain ${A}_{(1)}=\sum_{\mu=1}^n A_{\hat{\mu}}\,{e}^{\hat{\mu}}$.
Since $\partial \psi_k/\partial x_\mu=0$, we can use the functions $\{x_\mu,\psi_k\}$ as local coordinates.
In general, $A_{\hat{\mu}}$ may depend on $\psi_k$
and its dependence is determined by differential equation ${F}_{(2)}=d{A}_{(1)}$.
If we assume that components $A_{\hat{\mu}}$ are independent of $\psi_k$,
coordinates $\psi_k$ become Killing coordinates and 
the metric is explicitly given by
\begin{align}
 {g}
=& \sum_{\mu=1}^n \frac{U_\mu}{X_\mu}d x_\mu^2
   + \sum_{\mu=1}^n \frac{f_\mu^2X_\mu}{U_\mu}
    \Big(\sum_{k=0}^{n-1}A_\mu^{(k)}d\psi_k-{A}_{(1)}\Big)^2 ~,  \label{metevenA}
\end{align}
where
\begin{align}
 {A}_{(1)} = \frac{1}{\Phi}\sum_{\mu=1}^n \frac{N_\mu}{U_\mu}\,\sum_{k=0}^{n-1}A_\mu^{(k)}d\psi_k ~, \label{A1even}
\end{align}
$\Phi$ is given by \eqref{defH}, and torsion by \eqref{tsnevenA}.

\subsection{Type B: $M_{\mu\nu}=0$ case}
Next, we consider the case of $M_{\mu\nu}=0$ for all $\mu$ and $\nu$.
The torsion is fixed to be
\begin{align}
{T} = \sum_{\mu\neq\nu}\frac{2x_\mu\sqrt{Q_\nu}}{x_\mu^2-x_\nu^2}
         \,{e}^\mu\wedge{e}^{\hat{\mu}}\wedge{e}^{\hat{\nu}} ~.
\end{align}
Eq.\ (\ref{L1}) reduces to
\begin{align}
\partial_\nu L_\mu
= \frac{x_\nu L_\mu}{x_\mu^2-x_\nu^2}
  -\frac{2x_\mu x_\nu\sqrt{Q_\mu}}{(x_\mu^2-x_\nu^2)^2} ~,
\end{align}
which gives the solution
\begin{align}
L_\mu = -\sum_{\rho\neq\mu}\frac{x_\mu\sqrt{Q_\mu}}{x_\mu^2-x_\rho^2}
        +f_\mu\sqrt{Q_\mu} ~,
\end{align}
where $f_\mu$ are functions satisfying 
$\partial_\nu f_\mu={e}_{\hat{\nu}}(f_\mu)=0$ for $\mu\neq\nu$.

Let us consider vector fields $\{{\epsilon}_{\hat{\mu}}\}$ defined by
${\epsilon}_{\hat{\mu}} = \sqrt{U_\mu/Y_\mu}\,{e}_{\hat{\mu}}$,
where $\partial_\nu Y_\mu={e}_{\hat{\nu}}(Y_\mu)=0$ for $\mu\neq\nu$.
If $f_\mu=\partial_\mu \ln \sqrt{Y_\mu}$, these vector fields satisfy 
$ [{\epsilon}_\mu,{\epsilon}_{\hat{\nu}}]
=[{\epsilon}_{\hat{\mu}},{\epsilon}_{\hat{\nu}}]=0$.
Since we already have $[{\epsilon}_\mu,{\epsilon}_\nu]=0$, 
we can introduce local coordinates $y_\mu$ that are independent to $x_\mu$,
i.e., ${\epsilon}_{\hat{\mu}} = \partial/\partial y_\mu$.
Thus we have
\begin{align}
 {e}_\mu = \sqrt{\frac{X_\mu}{U_\mu}}\,\frac{\partial}{\partial x_\mu} ~,~~~~~
 {e}_{\hat{\mu}} = \sqrt{\frac{Y_\mu}{U_\mu}}\,\frac{\partial}{\partial y_\mu} ~.
 \label{orthoMzero}
\end{align}
Therefore, the metric is given by
\begin{align}
 {g}
= \sum_{\mu=1}^n U_\mu \Bigg(\,\frac{dx_\mu^2}{X_\mu}
  +\frac{dy_\mu^2}{Y_\mu}\,\Bigg) ~, \label{Mzerometric}
\end{align}
where $X_\mu$ and $Y_\mu$ are functions depending 
on both coordinates $x_\mu$ and $y_\mu$;
$X_\mu=X_\mu(x_\mu,y_\mu)$ and $Y_\mu=Y_\mu(x_\mu,y_\mu)$.
Thus, we have explicitly constructed metrics in all even dimensions which admit the whole tower of Killing tensors \eqref
{formK} but in general possess no Killing fields.

\subsection{Type C: Mixed case}
The mixed type C is the most complicated. For simplicity, we consider only four-dimensional case.
Eq.\ (\ref{MK}) then reads
\begin{align}
M_{12}K_2 = 0 ~,~~~ M_{21}K_1 = 0 ~. \label{MKeven}
\end{align}
When we choose $K_2=0$ and $M_{21}=0$,
the equations to solve are
\begin{align}
& \partial_y L_1 = \frac{y L_1}{x^2-y^2}
  -\frac{2x y\sqrt{Q_1}}{(x^2-y^2)^2} ~,~~~~~
  \partial_x L_2 = \frac{x L_2}{y^2-x^2}
  -\frac{2y x\sqrt{Q_2}}{(y^2-x^2)^2} ~, \nonumber\\
& \partial_y M_{12} = \Bigg(\frac{2y}{x^2-y^2}
-\frac{L_2}{\sqrt{Q_2}}\Bigg)M_{12} ~. 
\end{align}
The solutions are
\begin{align}
& L_1 = -\frac{x\sqrt{Q_1}}{x^2-y^2}+f_1\sqrt{Q_1} ~,~~~
  L_2 = -\frac{y\sqrt{Q_2}}{y^2-x^2}+f_2\sqrt{Q_2} ~, \nonumber\\
& M_{12} = h\sqrt{Q_1}\exp\left(-\int f_2\, dy\right)  ~,~~~
  M_{21}=0 ~, \label{CevenLM}
\end{align}
where $f_1$, $f_2$ and $h$ are arbitrary functions 
satisfying $\partial_yf_1=0$, $\partial_xf_2=0$, and $\partial_yh=0$.
Assuming $f_1=f_1(x)$, $f_2=f_2(y)$ and $h=h(x)$,
we have commutation relations
\begin{align}
& [{e}_1,{e}_2] 
= -\frac{y\sqrt{Q_2}}{x^2-y^2}\,{e}_1-\frac{x\sqrt{Q_1}}{x^2-y^2}\,{e}_2 ~, \nonumber\\
& [{e}_1,{e}_{\hat{1}}]
= K_1\,{e}_1+L_1\,{e}_{\hat{1}}+M_{12}\,{e}_{\hat{2}} ~,~~~
  [{e}_2,{e}_{\hat{2}}]
= L_2\,{e}_{\hat{2}} ~, \nonumber\\
& [{e}_1,{e}_{\hat{2}}]
= -\frac{x\sqrt{Q_1}}{x^2-y^2}\,{e}_{\hat{2}} ~,~~~
  [{e}_2,{e}_{\hat{1}}]
= -\frac{y\sqrt{Q_2}}{y^2-x^2}\,{e}_{\hat{1}} ~, \nonumber\\
& [{e}_{\hat{1}},{e}_{\hat{2}}] =0 ~.
\end{align}
It can be shown that the following vector fields $\{\epsilon_\mu,\hat{{\epsilon}}_\mu\}$
are mutually commuting:
\begin{align}
& \epsilon_1 = \frac{e_1}{\sqrt{Q_1}} ~,~~~
  \epsilon_2 = \frac{e_2}{\sqrt{Q_2}} ~, \nonumber\\
& \hat{{\epsilon}}_1 = E_1{}^1\,{e}_{\hat{1}} + E_1{}^2\,{e}_{\hat{2}} ~,~~~
 \hat{{\epsilon}}_2 = E_2{}^1\,{e}_{\hat{1}} + E_2{}^2\,{e}_{\hat{2}} ~,
\end{align}
where 
$\hat{{\epsilon}}_1=\partial/\partial u^1$, 
$\hat{{\epsilon}}_2=\partial/\partial u^2$ and
\begin{align}
& E_i{}^1 = \sqrt{x^2-y^2}\,a_i(u)\,\Psi_1 (x) ~, \nonumber\\
& E_i{}^2 = \sqrt{x^2-y^2}\Big(-a_i(u)\Xi_1(x)+b_i(u)\Big)\Psi_2(y) ~.
\end{align}
The functions $\Psi_1(x)$, $\Psi_2(y)$ and $\Xi_1(x)$ are given by
\begin{align}
& \Psi_1(x) = \exp\Big(-\int f_1(x)\, dx\Big) ~,~~~~~ 
  \Psi_2(y) = \exp\Big(-\int f_2(y)\, dy\Big) ~, \nonumber\\
& \Xi_1(x) = \int \Big(h(x)\exp\Big(-\int f_1(x)\, dx\Big)\Big)dx ~,
\end{align}
and $a_i(u)$ and $b_i(u)$ must satisfy
\begin{align}
 \frac{\partial a_2}{\partial u^1}-\frac{\partial a_1}{\partial u^2} = 0 ~,~~~~~
 \frac{\partial b_2}{\partial u^1}-\frac{\partial b_1}{\partial u^2} = 0 ~.
\end{align}
Moreover, in order that vector fields $\hat{{\epsilon}}_1$ and $\hat{{\epsilon}}_2$ 
are linear independent, it must be satisfied that $a_1b_2-a_2b_1\neq0$.
Then we obtain a local form of the metric
\begin{align}
 g = (x^2-y^2)\Bigg[\frac{dx^2}{X(x,\psi_1)}-\frac{dy^2}{Y(y)}
      +\Psi_1(x)^2d\psi_1^2+\Psi_2(y)^2\Big(-\Xi_1(x) d\psi_1 +d\psi_2\Big)^2\Bigg] ~,
\end{align}
where
\begin{align}
 d\psi_1 = a_1\,du^1+a_2\,du^2 ~,~~~~~
 d\psi_2 = b_1\,du^1+b_2\,du^2 ~.
\end{align}
We emphasize that $X$ allows dependence of $x$ and $\psi_1$, 
while $Y$ is a function depending only on $y$,
which comes from $K_1\neq0$ and $K_2=0$.

\section{Local metrics in odd dimensions}
In this section, we shall discuss local forms of metrics admitting a generalized PKY tensor in odd dimensions.
We shall proceed in a fashion similar to our approach in even dimensions.

\subsection{Connection and commutators}
\begin{lemma}
In odd dimensions, the connection 1-forms with torsion must have the following form:
\begin{align}\label{41}
{\omega}^T{}^\mu{}_\nu
=& -\frac{x_\nu\sqrt{Q_\nu}}{x_\mu^2-x_\nu^2}\,{e}^\mu
   -\frac{x_\mu\sqrt{Q_\mu}}{x_\mu^2-x_\nu^2}\,{e}^\nu ~,~~~~~(\mu\neq\nu) \nonumber\\
{\omega}^T{}^\mu{}_{\hat{\mu}}
=& -\frac{1}{\sqrt{Q_\mu}}\Bigg(
   \sum_{\nu\neq\mu}\frac{x_\mu Q_\nu}{x_\mu^2-x_\nu^2}
   +\frac{Q_0}{x_\mu} 
   \Bigg)\,{e}^{\hat{\mu}} +\sum_{\nu\neq\mu}
   \frac{x_\mu\sqrt{Q_\nu}}{x_\mu^2-x_\nu^2}\,{e}^{\hat{\nu}}\nonumber \\
 & +\frac{\sqrt{Q_0}}{x_\mu}\,{e}^0
   +\sum_a\frac{\kappa_a{}^\mu}{\sqrt{Q_\mu}}\,{e}^a ~, \nonumber\\
{\omega}^T{}^\mu{}_{\hat{\nu}}
=& \frac{x_\mu\sqrt{Q_\nu}}{x_\mu^2-x_\nu^2}\,{e}^{\hat{\mu}} 
    -\frac{x_\mu\sqrt{Q_\mu}}{x_\mu^2-x_\nu^2}\,{e}^{\hat{\nu}} ~,~~~~~(\mu\neq\nu) \\
{\omega}^T{}^{\hat{\mu}}{}_{\hat{\nu}}
=& -\frac{x_\mu\sqrt{Q_\nu}}{x_\mu^2-x_\nu^2}\,{e}^\mu
    -\frac{x_\nu\sqrt{Q_\mu}}{x_\mu^2-x_\nu^2}\,{e}^\nu ~,~~~~~(\mu\neq\nu) \nonumber\\
{\omega}^T{}^\mu{}_0
=& \frac{\sqrt{Q_0}}{x_\mu}\,{e}^{\hat{\mu}}
   -\frac{\sqrt{Q_\mu}}{x_\mu}\,{e}^0 ~, \nonumber\\
{\omega}^T{}^{\hat{\mu}}{}_0
=& -\frac{\sqrt{Q_0}}{x_\mu}\,{e}^\mu ~, \nonumber
\end{align}
where, as before, $\kappa_a{}^b$ is defined by \eqref{kappadef}.
\end{lemma}
To collect more information about $\kappa_a{}^b$, we differentiate (\ref{formxi}) using (\ref{41}), and obtain 
\begin{align}
& \kappa_{\mu\hat{\mu}}={e}_\mu(\sqrt{Q_\mu})
  -\sum_{\nu\neq\mu}\frac{x_\mu Q_\nu}{x_\mu^2-x_\nu^2}-\frac{Q_0}{x_\mu} ~, \\
& \kappa_{\mu\hat{\nu}}={e}_\mu(\sqrt{Q_\nu})
  +\frac{x_\mu\sqrt{Q_\mu}\sqrt{Q_\nu}}{x_\mu^2-x_\nu^2} ~,~~~(\mu\neq\nu) ~,~~~~~
  \kappa_{\mu 0}={e}_\mu(\sqrt{Q_0})
  +\frac{\sqrt{Q_\mu}\sqrt{Q_0}}{x_\mu} ~, \\
& \kappa_{\hat{\mu}\hat{\mu}}{}={e}_{\hat{\mu}}(\sqrt{Q_\mu}) ~, \label{4.9} \\
& \kappa_{\hat{\mu}\hat{\nu}}={e}_{\hat{\mu}}(\sqrt{Q_\nu}) ~,~~~(\mu\neq\nu) ~,~~~~~
  \kappa_{0\hat{\mu}}={e}_0(\sqrt{Q_\mu}) ~, \\
& \kappa_{\hat{\mu}0}={e}_{\hat{\mu}}(\sqrt{Q_0}) ~,~~~~~
  \kappa_{00}={e}_0(\sqrt{Q_0}) ~.
\end{align}
By using the integrability condition for the PKY tensor
we find that
\begin{align}
& \kappa_{\mu\mu}+\kappa_{\hat{\mu}\hat{\mu}}=0 ~, \label{4.12} \\
& \kappa_{\mu\nu} = \kappa_{\mu\hat{\nu}}=\kappa_{\hat{\mu}\hat{\nu}}=0 ~,~~~(\mu\neq\nu) ~,~~~~~
  \kappa_{0\hat{\mu}} = 0 ~, \label{4.13} \\
& \kappa_{\hat{\mu}\nu} = -\sqrt{Q_\nu}T_{\hat{\mu}\nu\hat{\nu}} ~,~~~ (\mu\neq\nu) ~,~~~~~
  \kappa_{0\mu} = -\sqrt{Q_\mu}T_{\mu\hat{\mu}0} ~,
\end{align}
and obtain
\begin{align}
& T_{\mu\nu\hat{\nu}}=T_{\mu\nu 0}=T_{\mu\hat{\nu}0}=0 ~,~~~ (\mu\neq\nu) ~, \\
& T_{\mu\nu\rho}=T_{\mu\nu\hat{\rho}}=T_{\mu\hat{\nu}\hat{\rho}}=0 ~.~~~(\text{$\mu,\nu,\rho$ all different})
\end{align}
From the connection 1-forms (\ref{41}) together with (\ref{4.13})
one can evaluate covariant derivatives $\nabla^T_{e_a}{e}_b$,
which are summarized in App.\ C.
Using these formulae and Eq (\ref{eq3-11}) we can directly confirm that 
(\ref{formH}) satisfies the PKY equation. 

From Eq. (\ref{3.16}), we have the commutation relations
\begin{align}
\left[{e}_\mu,{e}_\nu \right] 
=& -\frac{x_\nu\sqrt{Q_\nu}}{x_\mu^2-x_\nu^2}\,{e}_\mu
    -\frac{x_\mu\sqrt{Q_\mu}}{x_\mu^2-x_\nu^2}\,{e}_\nu ~,~~~(\mu\neq\nu) \label{comod1} \\
\left[{e}_\mu,{e}_{\hat{\mu}}\right]
=& K_\mu\,{e}_\mu
  +L_\mu\,{e}_{\hat{\mu}}
  +\sum_{\rho\neq\mu}M_{\mu\rho}\,{e}_{\hat{\rho}}
  +J_\mu\,{e}_0 ~, \\
\left[{e}_\mu,{e}_{\hat{\nu}}\right]
=&-\frac{x_\mu\sqrt{Q_\mu}}{x_\mu^2-x_\nu^2}\,{e}_{\hat{\nu}} ~,~~~(\mu\neq\nu) \\
\left[{e}_{\hat{\mu}},{e}_{\hat{\nu}}\right]
=& -\sum_{\rho\neq\mu,\nu}T_{\hat{\mu}\hat{\nu}\hat{\rho}}\,{e}_{\hat{\rho}}
    -T_{\hat{\mu}\hat{\nu}0}\,{e}_0 ~,~~~(\mu\neq\nu) \\
\left[{e}_\mu,{e}_0\right]
=& -\frac{\sqrt{Q_\mu}}{x_\mu}\,{e}_0 ~, \\
\left[{e}_{\hat{\mu}},{e}_0\right]
=& \sum_{\nu\neq\mu}T_{\hat{\mu}\hat{\nu}0}\,{e}_{\hat{\nu}} ~, \label{comod4}
\end{align}
where
\begin{align}
& K_\mu \equiv \frac{\kappa_\mu{}^\mu}{\sqrt{Q_\mu}} ~,~~~
  L_\mu \equiv -\frac{1}{\sqrt{Q_\mu}}\Bigg(\sum_{\rho\neq\mu}
  \frac{x_\mu Q_\rho}{x_\mu^2-x_\rho^2}+\frac{Q_0}{x_\mu}-\kappa_{\hat{\mu}}{}^\mu\Bigg) ~, \nonumber\\
& M_{\mu\nu} \equiv \frac{2x_\mu\sqrt{Q_\nu}}{x_\mu^2-x_\nu^2}
  -T_{\mu\hat{\mu}\hat{\nu}} ~~(\mu\neq\nu) ~,~~~
  J_\mu \equiv\frac{2\sqrt{Q_0}}{x_\mu}-T_{\mu\hat{\mu}0} ~. \label{defKLMJ}
\end{align}
Especially, from (\ref{4.9}) and (\ref{4.12}) we again obtain (\ref{Kmu}).
Moreover, we have
\begin{align}
{e}_\mu = \sqrt{Q_\mu}\frac{\partial}{\partial x_\mu} ~,
\end{align}
where $Q_\mu$ takes the form (\ref{Qmu}) with functions $X_\mu$
satisfying ${e}_\nu(X_\mu)={e}_{\hat{\nu}}(X_\mu)={e}_0(X_\mu)=0$.
The Jacobi identities require $T_{\hat{\mu}\hat{\nu}\hat{\rho}}=0$ and $T_{\hat{\mu}\hat{\nu}0}=0$,
which leads to Lemma 2.1.
Unknown functions $K_\mu$, $L_\mu$ and $M_{\mu\nu}$ obey the same equations\footnote{
Note that functions $L_\mu$ in odd dimensions are different from 
those in even dimensions, though we use the same symbols $L_\mu$, 
cf. (\ref{KLMdef}).} 
(\ref{MK})--(\ref{originaleven}) and in addition
\begin{align}
 \partial_\nu J_\mu 
  =  \Big(\frac{2x_\nu}{x_\mu^2-x_\nu^2}
        +\frac{1}{x_\nu}\Big)J_\mu-\frac{M_{\mu\nu}J_\nu}{\sqrt{Q_\nu}} ~,~~~(\mu\neq\nu) \label{J21}
\end{align}
and
\begin{align}
& {e}_0(K_\mu)=0 ~,~~~~~
  {e}_0(L_\mu)=0 ~,~~~~~
  {e}_0(M_{\mu\nu})=0 ~~~(\mu\neq\nu)~, \label{J22}\\
& {e}_{\hat{\nu}}(J_\mu) =0 ~~~(\mu\neq\nu) ~,~~~~~
  {e}_0(J_\mu)=0 ~. \label{J23}
\end{align}
We already know a solution for functions $K_\mu$, $L_\mu$ and $M_{\mu\nu}$ 
because they obey the same equations as in even dimensions; the dependence of ${e}_0$ is determined by (\ref{J22}).
So again, we obtain three classes of solutions for $K_\mu$, $L_\mu$ and $M_{\mu\nu}$.
On the other hand, having one more function $J_\mu$ in the case of odd dimensions, 
we still have to solve differential equations (\ref{J21}) and (\ref{J23}).
The torsion always includes one arbitrary function $Q_0$, cf.\ (\ref{defKLMJ}).
The frame is determined from (\ref{3.32}), (\ref{3.33}), and 
\begin{align}
d{e}^0
=& -\sum_{\mu=1}^nJ_\mu\,{e}^\mu\wedge{e}^{\hat{\mu}}
   +\sum_{\mu=1}^n\frac{\sqrt{Q_\mu}}{x_\mu}\,{e}^\mu\wedge{e}^0 ~. \label{de0}
\end{align}

\subsection{Type A: $K_\mu=0$ case}
Taking $K_\mu=0$ for all $\mu$, 
we have the solutions (\ref{sol31}) and (\ref{sol32}).
Substituting (\ref{sol32}) into (\ref{J21}), we obtain
\begin{align}
\partial_\nu J_\mu
=& \Big(\frac{2x_\nu}{x_\mu^2-x_\nu^2}+\frac{1}{x_\nu}\Big)J_\mu
   -\frac{f_\nu}{f_\mu}\Big(\frac{2x_\mu}{x_\mu^2-x_\nu^2}+\partial_\mu \ln \Phi\Big)J_\nu ~.~~~(\mu\neq\nu) \label{eq418}
\end{align}
We find $J_\mu=k_1J_\mu^{(1)}+k_2J_\mu^{(2)}$ 
as a linear combination of two solutions
\begin{align}
 J_\mu^{(1)} = \frac{1}{f_\mu \prod_{\rho=1}^n x_\rho}\Big(\frac{2}{x_\mu}
     +\partial_\mu\ln\Phi\Big) ~,~~~
 J_\mu^{(2)} = \frac{1}{f_\mu}(\partial_\mu\ln\Phi)\prod_{\mu=1}^n x_\mu ~,
\end{align}
where $k_1$ and $k_2$ are arbitrary constants and $\Phi$ is given by \eqref{defH}.
By parallel calculations to even dimensions, leading to metrics \eqref{metevenA}, we obtain the following solution: 
\begin{align}
 {g}
=& \sum_{\mu=1}^n \frac{U_\mu}{X_\mu}d x_\mu^2
   + \sum_{\mu=1}^n \frac{f_\mu^2X_\mu}{U_\mu}
    \Bigg(\sum_{k=0}^{n-1}A_\mu^{(k)}d\psi_k-{A}_{(1)}\Bigg)^2 \nonumber\\
 & +\Bigg(\frac{k_1}{\prod_{\rho=1}^n x_\rho}\Big(\sum_{k=0}^nA^{(k)}
    d\psi_k-{A}_{(1)}\Big)
   +k_2\Big(\prod_{\rho=1}^n x_\rho\Big)\Big(d\psi_n-{A}_{(1)}\Big)\Bigg)^2 ~,
 \label{metodda}
\end{align}
where ${A}_{(1)}$ is given by (\ref{A1even}), and the torsion takes the form
\begin{align}
{T}
  =& \sum_{\mu\neq\nu}\Bigg[\frac{2x_\mu}{x_\mu^2-x_\nu^2}
    -\frac{f_\nu}{f_\mu}\Big(\frac{2x_\mu}{x_\mu^2-x_\nu^2}
    +\partial_\mu\ln\Phi\Big)\Bigg]\sqrt{Q_\nu}
    \,{e}^\mu\wedge {e}^{\hat{\mu}}\wedge{e}^{\hat{\nu}} \nonumber\\
   & +\sum_{\mu=1}^n \Bigg[\frac{2\sqrt{Q_0}}{x_\mu}
      -\frac{k_1}{f_\mu \prod_{\rho=1}^n x_\rho}
       \Big(\frac{2}{x_\mu}+\partial_\mu\ln\Phi\Big)
      -\frac{k_2}{f_\mu}(\partial_\mu\ln\Phi)\prod_{\rho=1}^n x_\rho\Bigg]
     \,{e}^\mu\wedge{e}^{\hat{\mu}}\wedge{e}^0 ~. \label{tsnodda}
\end{align}

Specifically, we consider the case of $T_{\mu\hat{\mu}\hat{\nu}}=0$ for $\mu\neq\nu$. In this case, the PKY tensor becomes both $d$-closed and $d^T$-closed, 
cf. \eqref{dh2}.
Then we have
\begin{align}
 L_\mu = -\partial_\mu\sqrt{Q_\mu} ~,~~~
 M_{\mu\nu} = \frac{2x_\mu\sqrt{Q_\nu}}{x_\mu^2-x_\nu^2} ~,
\end{align}
and Eq.\ (\ref{J21}) has the solution
\begin{equation}
J_\mu=k\Big(\prod_{\rho=1}^n x_\rho\Big)(\partial_\mu\Phi) 
\end{equation}
with an arbitrary constants $k$.
The torsion takes the form
\begin{align}
{T}
  = \sum_{\mu=1}^n \Bigg[\frac{2\sqrt{Q_0}}{x_\mu}
      - k \Big(\prod_{\rho=1}^n x_\rho\Big)(\partial_\mu\Phi)\Bigg]
     \,{e}^\mu\wedge{e}^{\hat{\mu}}\wedge{e}^0 ~, \label{tsnoddc}
\end{align}
and the corresponding metric is given by
\begin{align}
 {g}
=& \sum_{\mu=1}^n \frac{U_\mu}{X_\mu} dx_\mu^2+\sum_{\mu=1}^n \frac{X_\mu}{U_\mu} 
   \Bigg( \sum_{k=0}^{n-1}A_\mu^{(k)}d\psi_k \Bigg)^2
   +k^2\Big(\prod_{\rho=1}^nx_\rho^2\Big)\Big( d\psi_n
 -{B}_{(1)}\Big)^2 ~, \label{metoddA2}
\end{align}
where ${B}_{(1)}$ is defined by
\begin{equation}
B_{(1)}= \sum_{\mu=1}^n \frac{N_\mu}{U_\mu} \sum_{k=0}^{n-1}A_\mu^{(k)} d\psi_k ~. 
\label{B1}
\end{equation}
Both metrics \eqref{metodda} and \eqref{metoddA2} provide an ansatz for supergravity solutions and will be exploited in the next section.

\subsection{Type B: $M_{\mu\nu}=0$ case}
When we take $M_{\mu\nu}=0$ for all $\mu$ and $\nu$, then
\begin{align}
T= \sum_{\mu\neq \nu}\frac{2x_\mu\sqrt{Q_\nu}}{x_\mu^2-x_\nu^2} e^\mu\wedge e^{\hat \mu}\wedge e^{\hat \nu}~.
\end{align}
Eq. (\ref{J21}) reduces to
\begin{align}
 \partial_\nu J_\mu = 
  \Big(\frac{2x_\nu}{x_\mu^2-x_\nu^2}+\frac{1}{x_\nu}\Big)J_\mu ~.
\end{align}
In the same manner as in even dimensions we obtain (\ref{orthoMzero}), 
which implies that $\mu\hat{\mu}$-components of $de^0$ vanish.
Hence, from Eq. (\ref{de0}) we have $J_\mu=0$ and
\begin{align}
 {g}
=& \sum_{\mu=1}^nU_\mu\Big(\frac{dx_\mu^2}{X_\mu}
  +\frac{dy_\mu^2}{Y_\mu}\Big)
   +\Big(\prod_{\mu=1}^nx_\mu^2\Big)\,dz^2 ~,
\end{align}
where $X_\mu$ and $Y_\mu$ are functions depending 
on both coordinates $x_\mu$ and $y_\mu$;
$X_\mu=X_\mu(x_\mu,y_\mu)$ and $Y_\mu=Y_\mu(x_\mu,y_\mu)$.

\subsection{Type C: Mixed case}
For simplicity, we consider five-dimensional case.
Since Eq. (\ref{MK}) must hold in the case of odd dimensions, 
we take $K_2=0$ and $M_{21}=0$.
Then one finds the solutions (\ref{CevenLM}) for $L_\mu$ and $M_{\mu\nu}$,
and obtains the partial differential equations for $J_\mu$,
\begin{align}
 \frac{\partial J_1}{\partial y}=\Big(\frac{2y}{x^2-y^2}+\frac{1}{y}\Big)J_1
 -\frac{M_{12}J_2}{\sqrt{Q_2}} ~,~~~~~
 \frac{\partial J_2}{\partial x}=\Big(\frac{2x}{y^2-x^2}+\frac{1}{x}\Big)J_2 ~.
\end{align}
The solutions are
\begin{align}
 J_1 = \frac{yS_1\sqrt{Q_1}}{\sqrt{x^2-y^2}} ~,~~~~~
 J_2 = \frac{xS_2\sqrt{Q_2}}{\sqrt{x^2-y^2}} ~,
\end{align}
where $\partial S_2/\partial x=0$ and
\begin{align}
 S_1 = -x h \int \frac{S_2 \Psi_2}{y}\, dy
\end{align}
with $\Psi_1$, $\Psi_2$ and $h$ defined in Sec.~3.5.
One also obtains commuting vector fields $\hat{{\epsilon}}_i = E_i{}^1\,{e}_{\hat{1}} 
+ E_i{}^2\,{e}_{\hat{2}} + E_i{}^0\,{e}_0$ ($i=0,1,2$), where 
$\hat{{\epsilon}}_0=\partial/\partial u^0$,
$\hat{{\epsilon}}_1=\partial/\partial u^1$,
$\hat{{\epsilon}}_2=\partial/\partial u^2$ and
\begin{align}
& E_i{}^1 = \sqrt{x^2-y^2}\,a_i(u)\,\Psi_1(x) ~,~~~
  E_i{}^2 = \sqrt{x^2-y^2}\,\Big(-a_i(u) \Xi_1(x) + b_i(u) \Big)\,\Psi_2(y) ~, \nonumber\\
& E_i{}^0 = xy\Bigg[\Big(a_i(u)\Xi_1(x)-b_i(u)\Big)\Xi_2(y)+c_i(u)\Bigg] ~.
\end{align}
The functions $\Xi_1(x)$ and $\Xi_2(y)$ are
\begin{align}
 \Xi_1(x) = \int \Big(h \exp\Big(-\int f_1 dx\Big)\Big)dx ~,~~~
 \Xi_2(y) = \int\frac{s_2\Psi_2}{y}\,dy ~,
\end{align}
and ${a}=(a_i)$, ${b}=(b_i)$ and ${c}=(c_i)$ $(i=0,1,2)$ must satisfy
${\nabla}\times {a} = 0$, ${\nabla}\times {b} = 0$ and ${\nabla}\times {c} = 0$,
where ${\nabla}=(\partial/\partial u^0,\partial/\partial u^1,\partial/\partial u^2)$.
Moreover, in order that the vector fields $\hat{{\epsilon}}_i$ are linearly independent,
${a}$, ${b}$ and ${c}$ are also linearly independent.
Thus the five-dimensional metric of type C takes a local form
\begin{align}
 {g}
=& (x^2-y^2)\Bigg(\frac{dx^2}{X(x,\psi_1)}-\frac{dy^2}{Y(y)}
   +\Psi_1(x)^2d\psi_1^2+\Psi_2(y)^2\Big(-\Xi_1(x)d\psi_1+d\psi_2\Big)^2\Bigg) \nonumber\\
 & +x^2y^2 \Bigg(d\psi_0-\Xi_2(y)\Big(-\Xi_1(x)d\psi_1+d\psi_2\Big)\Bigg)^2 ~,
\end{align}
where $d\psi_1 = \sum_i a_i\,du^i$, $d\psi_2=\sum_ib_i\,du^i$ and $d\psi_0=\sum_ic_i\,du^i$.
It should be emphasized that $X$ depends on $x$ and $\psi_1$ 
and $Y$ depends only on $y$.

\section{Physical examples}
In this section we shall illustrate 
how the results described in Sec. 3 and Sec. 4 can be applied in
concrete supergravity theories. 
In arbitrary even dimensions we present new
examples of K\"ahler with torsion (KT) metrics. 
These are obtained by slightly modifying
the ansatz of higher dimensional charged Kerr-NUT black hole 
metrics.\footnote{The deformations of Calabi--Yau manifolds 
as supersymmetric solutions to abelian heterotic supergravity are discussed in
 \cite{Martelli:2010}. Although the method is different, our KT examples are
closely related to their deformations.}

\subsection{Solutions of heterotic supergravity}
We consider the abelian heterotic supergravity, 
which is obtained as a low-energy effective theory of heterotic string theory.
The action consists of a metric ${g}$, scalar field $\phi$, 
$U(1)$ potential ${A}$ and 2-form potential ${B}$,
\begin{align}
 S = \int e^{\phi}\Big(*{\cal R}+*d\phi\wedge d\phi
     -*{F}\wedge {F}-\frac{1}{2}*{H}\wedge {H}\Big)
\end{align}
where ${F}={dA}$ and ${H}={dB}-{A}\wedge{dA}$.
The equations of motion are 
\begin{align}
& R_{ab}-\nabla_a \nabla_b \phi-F_{a}^{~c}F_{bc}-\frac{1}{4}H_{a}^{~cd}H_{bcd}=0 ~, \label{eom1}\\
& d(e^{\phi} *{F})=e^{\phi} *{H}\wedge {F} ~, \label{eom2}\\
& d(e^{\phi} *{H})=0 ~, \label{eom3}\\
& (\nabla\phi)^2+2 \nabla^2\phi+\frac{1}{2}F_{ab}F^{ab}+\frac{1}{12}H_{abc}H^{abc}-R = 0 ~. \label{eom4}
\end{align}

We investigate solutions whose metrics take the form of type A, i.e.,
(\ref{metevenA}) in even dimensions and (\ref{metodda}) in odd dimensions.
The metric in $2n+\varepsilon$ dimensions ($\varepsilon=0$ or $1$)
includes unknown functions $f_\mu$, $N_\mu$ and $X_\mu$ ($\mu=1,\cdots,n$) 
which depend on one variable $x_\mu$ only.
In particular, we find solutions in two cases of $f_\mu=1$ and $f_\mu=2x_\mu$.

\subsubsection{Charged Kerr-NUT black hole metrics}
In the case of $f_\mu=1$ for all $\mu$, 
we use ${F}_{(2)}=d{A}_{(1)}$, (\ref{2-form}), 
as ${F}$ and further identify ${H}$ with the torsion 3-form ${T}$, 
(\ref{tsnevenA}) or (\ref{tsnodda}).
In odd dimensions, we have one more arbitrary function $Q_0$ 
in the torsion ${T}={H}$.
The equations of motion (\ref{eom1})--(\ref{eom4})
give the solution
\begin{align}
& X_\mu = \sum_{k=0}^{n-1}c_kx_\mu^{2k}+m_\mu x_\mu^{1-\varepsilon}
        +\varepsilon \frac{(-1)^nc}{x_\mu^2} ~, \\
& N_\mu = \sum_{k=0}^{n-1}b_k x_\mu^{2k}+a m_\mu x_\mu^{1-\varepsilon} ~. 
\end{align}
In odd dimensions, the function $Q_0$ 
is determined as $\sqrt{Q_0}=\sqrt{c}/\prod_{\rho=1}^nx_\rho$.
This solution includes free parameters $m_\mu$ ($\mu=1,\cdots,n$), 
$a$, $b_k$ and $c_k$ ($k=0,\cdots,n-1$),
$c$ and $q$ with a relation $1+b_{n-1}=a c_{n-1}+a q^2$.
If we take $b_k=0$ for $k=0,\cdots,n-1$, 
the solution reproduces the charged Kerr-NUT black hole solution 
\cite{Sen:1992,Cvetic2:1996,Chow2:2010}. 
Then the metric and the fields are given by (in odd dimensions we have chosen $k_1=\sqrt{c}$ and $k_2=0$) 
\begin{align} \label{TCKY}
 g 
=& \sum_{\mu=1}^n \frac{U_\mu}{X_\mu}dx_\mu^2
   +\sum_{\mu=1}^n \frac{X_\mu}{U_\mu} 
    \left( \sum_{k=0}^{n-1} A^{(k)}_\mu d\psi_k
     -{A}_{(1)}\right)^2\nonumber\\
 & + \frac{\varepsilon c}{\prod_{\rho=1}^nx_\rho^2}\left( \sum_{k=0}^n A^{(k)} d\psi_k
     -{A}_{(1)}\right)^2 ~, \\
 F =& q \sum_{\mu=1}^n \partial_\mu \ln\Phi\,{e}^\mu\wedge{e}^{\hat{\mu}} ~, 
  \label{F11}\\
 H =& -\left(\sum_{\mu=1}^n \partial_\mu \ln\Phi
         \, {e}^\mu\wedge{e}^{\hat{\mu}}\right)
         \wedge \left(\sum_{\nu=1}^n \sqrt{\frac{X_\nu}{U_\nu}}\,{e}^{\hat{\nu}}
               + \frac{\varepsilon\sqrt{c}}{\prod_{\rho=1}^nx_\rho}\,{e}^0\right) ~,
  \label{H11} \\
 \phi=& \ln\Phi ~,
\end{align}
where 
\begin{equation}
\Phi = 1 + b_{n-1} + a \sum_{\mu=1}^n \frac{m_\mu x_\mu^{1-\varepsilon}}{U_\mu} ~,
\end{equation}
and ${A}_{(1)}$ is given by (\ref{A1even}). 
Properties of these solutions related to the hidden symmetries have been 
studied in \cite{Chow2:2010, Houri:2010fr}.
In even dimensions, the metric (\ref{TCKY}) is hermitian 
for complex structures ${J}_{\epsilon}$ defined in App.\ B,
and hence the charged Kerr-NUT metric has multi-hermitian structure\footnote{
When the torsion is absent, 
the multi-hermitian structure of the Kerr-NUT-(A)dS metrics was discussed in \cite{Mason:2010}.}.
The corresponding KT structure is given by the Bismut torsion \cite{Bismut:1989} 
\begin{align}
 B_\epsilon = \sum_{\mu=1}^n\sum_{\nu\neq\mu}
  \Bigg(\frac{2(\epsilon_\mu\epsilon_\nu x_\nu-x_\mu)}{x_\mu^2-x_\nu^2}
               -\partial_\mu \ln \Phi\Bigg)\sqrt{\frac{X_\nu}{U_\nu}}
  \,e^\mu\wedge e^{\hat{\mu}}\wedge e^{\hat{\nu}} ~,
\end{align}
with $\epsilon_\mu=\pm 1$.

\subsubsection{Calabi--Yau with torsion metrics}
We consider the even dimensional metric (\ref{metevenA}) with $f_\mu=2x_\mu$ for $\mu=1,\cdots,n$,
\begin{align}
 g 
= \sum_{\mu=1}^n \frac{U_\mu}{X_\mu}dx_\mu^2
   +\sum_{\mu=1}^n \frac{4x_\mu^2X_\mu}{U_\mu} 
    \left( \sum_{k=0}^{n-1} A^{(k)}_\mu d\psi_k
     -{A}_{(1)}\right)^2 ~. \label{metKT}
\end{align}
For the metric (\ref{metKT}) and the complex structures $J_\epsilon$ defined in App.\ B, 
the Bismut torsion is given by
\begin{align}
{B}_\epsilon
  = \sum_{\mu\neq\nu}\Big(
  \frac{2(\epsilon_\mu\epsilon_\nu-1)x_\nu}{x_\mu^2-x_\nu^2}
  -\frac{x_\nu \big(\partial_\mu\ln\Phi\big)}{x_\mu}\Big)\sqrt{\frac{X_\nu}{U_\nu}}
    \,{e}^\mu\wedge {e}^{\hat{\mu}}\wedge{e}^{\hat{\nu}}
\end{align}
with $\epsilon_\mu=\pm 1$. 
In turn, instead of indentifying the 3-form field strength $H$ with the torsion $T$
associated with PKY tensor,
we use the Bismut torsion $B_{\epsilon}$ 
with all $\epsilon_\mu$ equal, $\epsilon_1=\epsilon_2=\cdots=\epsilon_n=\pm1$.
And we take the Maxwell field $F$ as (\ref{2-form}).
The equations of motion give the solution
\begin{align}
& X_\mu = \frac{1}{4x_\mu^2}\Big(\sum_{k=1}^n c_kx_\mu^{2k}+m_\mu\Big) ~, \\
& N_\mu = \sum_{k=1}^n b_k x_\mu^{2k}+a m_\mu
\end{align}
with $b_n=ac_n$. 
This solution includes free parameters $m_\mu$ ($\mu=1,\cdots,n$), 
$a$, $b_k$ and $c_k$ ($k=1,\cdots,n$) and $q$ with a relation $1+b_{n-1}=a c_{n-1}+a q^2$.
Then the fields are given by
\begin{align} 
 {F} =& q\sum_{\mu=1}^n\frac{\partial_\mu\ln\Phi}{2x_\mu}
         \,{e}^\mu\wedge{e}^{\hat{\mu}} ~, \\
 {H} =& -\sum_{\mu\neq\nu}\frac{x_\nu(\partial_\mu\ln\Phi)}{x_\mu}\sqrt{\frac{X_\nu}{U_\nu}}
    \,{e}^\mu\wedge {e}^{\hat{\mu}}\wedge{e}^{\hat{\nu}} ~, \\
 \phi =& \ln\Phi ~,
\end{align}
where
\begin{equation}
 \Phi = 1 + b_{n-1} + b_n \sum_{\mu=1}^nx_\mu^2+a\sum_{\mu=1}^n\frac{m_\mu}{U_\mu} ~.
\end{equation}

In particular, when we put $c_n=0$ and $\epsilon_1=\epsilon_2=\cdots=\epsilon_n=\pm1$,
the Ricci form ${\rho}^B$ associated with the Bismut connection \cite{Ivanov:2001,Friedrich:2002} 
\begin{align}
 {\rho}^B({X},{Y})=
\frac{1}{2}\sum_{a=1}^{2n}{\cal R}^B({X},{Y},{e}_a,{J}({e}_a))
\end{align}
vanishes. Therefore the metric becomes Calabi--Yau with torsion.

If we take $a=0$ and $b_k=0$ for all $k$ 
with $\epsilon_1=\epsilon_2=\cdots=\epsilon_n=\pm1$, 
the torsion ${B}_\epsilon$ vanishes 
and the metric reduces to the orthotoric K\"ahler metric 
fully studied in \cite{Apos:2006}.
In this case the torsion 3-form $T$ remains non-trivial,
\begin{align}
 {T} = \sum_{\mu\neq\nu}\frac{2}{x_\mu+x_\nu}\sqrt{\frac{X_\nu}{U_\nu}}
    \,{e}^\mu\wedge {e}^{\hat{\mu}}\wedge{e}^{\hat{\nu}} ~.
\end{align}
This means that the orthotoric K\"ahler manifold does not admit ordinary
closed conformal Killing--Yano tensors but possesses the PKY tensor with torsion.\footnote{
It is known that there exists a Hamiltonian 2-form 
which always produces a rank-2 conformal Killing--Yano tensor.
However, this is neither closed nor co-closed \cite{Apos:2006}.} 
It is this tensor which is responsible for separability of Laplacian in these spaces, cf., \cite{Chen:2007em}.

\subsection{Five-dimensional minimal supergravity black hole metrics}
We consider the five-dimensional minimal gauged supergravity.
The action is given by 
\begin{align}
S = \int *({\cal R}+\Lambda)-\frac{1}{2}{F}\wedge*{F}
        +\frac{1}{3\sqrt{3}}{F}\wedge{F}\wedge{A}
\end{align}
where ${F}$ is 2-form field strength of Maxwell field ${A}$, ${F}=d{A}$.
The equations of motion are
\begin{align}
& R_{ab} +\frac{\Lambda}{3}g_{ab} 
= \frac{1}{2}\Big(F_{ac}F_b{}^c-\frac{1}{6}g_{ab}F_{cd}F^{cd}\Big) ~, \\
& d*{F}-\frac{1}{\sqrt{3}}{F}\wedge{F} = 0 ~.
\end{align}
We investigate five-dimensional metrics written in the form (\ref{metoddA2}).
This metric includes unknown functions $N_\mu$ and $X_\mu$ ($\mu=1,2$) 
which depend on one variable $x_\mu$ only.
Following \cite{Kubiznak:2009a}, we identify ${F}$ with the torsion 3-form by ${T}=*{F}/\sqrt{3}$ 
.
The equations of motion give the solution
\begin{align}
 X_\mu =& c_2x_\mu^4+c_1x_\mu^2+m_\mu+\frac{k^2q_\mu^2}{x_\mu^2} ~, \\
 N_\mu =& b_1x_\mu^2+b_0+\frac{q_\mu}{x_\mu^2} ~,
\end{align}
where $c_2=\Lambda/30$ and  
$c_1$, $b_0$, $b_1$ $m_\mu$, $k$, $q_1$ and $q_2$ 
are free parameters\footnote{
The parameters $k$ is a pure imaginary constant.}.
The function $Q_0$ in (\ref{tsnoddc}) is determined as
\begin{align}
 \sqrt{Q_0} = \frac{k(q_2x^2-q_1y^2)}{xy(x^2-y^2)} ~.
\end{align}
The metric and the Maxwell field are given by
\ba
 {g}
&=& \sum_{\mu=1}^2\frac{U_\mu}{X_\mu}dx_\mu^2
    +\sum_{\mu=1}^2\frac{X_\mu}{U_\mu}\Bigg(\sum_{k=0}^1A_\mu^{(k)}d\psi_k\Bigg)^2 
    +k^2x^2y^2\Big(d\psi_2-{B}_{(1)}\Big)^2 ~, \label{CCLPmetric}\\
 {F}
&=& \frac{2\sqrt{3}k(q_1-q_2)}{(x^2-y^2)^2}
   \Big(x\,{e}^1\wedge{e}^{\hat{1}}-y\,{e}^2\wedge{e}^{\hat{2}}\Big) ~,
\ea
where ${B}_{(1)}$ is defined by (\ref{B1}).

This metric reproduces the rotating black hole metric discovered by Chong, Cvetic, L\"u, and Pope in \cite{Chong:2005b}.
In the paper \cite{Ahmedov:2009}, it was shown that 
(\ref{CCLPmetric}) is a unique metric admitting the $d$-closed PKY tensor with torsion subject to 
certain additional assumptions. Here we see how the demonstrated uniqueness fits into a general picture of classification of metrics admitting the PKY tensor with torsion.

\section{Conclusions}
In this paper we have classified spacetimes admitting a non-degenerate 
rank-2 $d^T\!$-closed generalized conformal Killing--Yano (PKY) tensor in all dimensions $D$.
This classification, apart from its own significance,  provides an alternative to various approaches to constructing new exact solutions.
In particular, the spacetimes obtained  provide an ansatz for exact solutions of various supergravities.
A remarkable property of these metrics is that
the PKY tensor generates a set of $n=[D/2]$ mutually commuting rank-2 Killing tensors. If a sufficient number of additional isometries is present (as is the case for the physical examples we discuss), this guarantees complete integrability of the geodesic equations as well and we expect furthermore the separability of scalar and Dirac equations.
The problem of classification has been reduced
to that of solving certain partial differential equations
(\ref{L1})--(\ref{M2}) in even dimensions and/or (\ref{J21}) in odd dimensions.
The solutions can be classified into three types (A, B and C); we have constructed the corresponding examples of metrics explicitly.

So far we have not been able to find a general solution 
to the partial differential equations obtained and hence complete classification remains an open issue.
However, we have demonstrated that our metrics cover many known solutions of various supergravities, such as
higher-dimensional Kerr--Sen black hole metrics, KT metrics 
and Calabi--Yau with torsion metrics 
in abelian heterotic supergravity, and 
the charged rotating black hole metric 
of five-dimensional minimal gauged supergravity.
{We believe that the KT and Calabi--Yau with torsion metrics constructed in this paper are new.}
Recently constructed black hole solutions 
of gauged supergravities in 4, 6 and 7 dimensions \cite{Chow:2008,Chow:2010,Chow:2011}
are also included in our metric.
When studying the physical significance of the metrics we obtain, we have concentrated on type A metrics.
It would be interesting to examine the physical meaning of type B and C metrics in the future. 

One possible generalization of the obtained classification would be to relax the assumption on the non-degeneracy of the PKY tensor. 
In the torsion-less case, this leads to much richer structure of spacetimes, while the full classification is still possible \cite{Houri:2008b, Houri:2009}. 
We also believe that the ansatz of higher rank GCKY tensors would lead to new families of more general solutions.
Since our analysis was local, it is desirable to obtain global description of the metrics as a future problem.
Global properties of conformal Killing--Yano tensors were
investigated by Semmelmann \cite{Semmelmann:2002}. He showed the existence
of Killing--Yano tensors on Sasakian, 3-Sasakian, nearly Kahler and weak
$G_2$-manifold. These geometries
are deeply related to supersymmetric compactifications and AdS/CFT
correspondence in string theories.
In App.~A we have discussed generalized Killing spinors according to
Semmelmann's argument. It is an interesting question whether the
presented method, or its generalizations, can provide a new construction
in the geometry with torsion.

\acknowledgments 
We wish to thank G.~W.~Gibbons and H.~S.~Reall for discussions and reading the manuscript.
T.H. and Y.Y. also would like to thank DAMTP, University of Cambridge, for the hospitality.
The work of T.H. is partially supported by the JSPS Institutional Program
for Young Researcher Overseas Visits ``Promoting international young
researchers in mathematics and mathematical sciences led by OCAMI'' and
the JSPS Strategic Young Researcher Overseas Visits Program for
Accelerating Brain Circulation ``Deeping and Evolution of Mathematics and
Physics, Building of International Network Hub based on OCAMI.''
The work of Y.Y. is supported by the Grant-in Aid for Scientific Research 
No.\ 21244003 and No.\ 23540317 from Japan Ministry of Education. 
C.W. is grateful to PIMS and NSERC for support.

\appendix

\section{Torsion Killing spinors}\label{twist}
In this appendix we establish the relation between generalized Killing--Yano tensors and various torsion Killing spinors by extending the work of Semmelmann \cite{Semmelmann:2002} and Cariglia 
\cite{Cariglia:2004}. 
To make calculations feasible we use the compact notations of \cite{HouriEtal:2010}. Namely, we identify the elements of Clifford algebra with differential forms and denote the Clifford product by juxtaposition. Namely, for a $1$-form $\alpha$ and $p$-form $\omega$ this reads 
\be\label{Cliff}
\alpha \omega = \alpha \wedge \omega + \alpha^\sharp \hook \omega\,,\quad 
\omega \alpha = (-1)^p \bigl(\alpha \wedge \omega - \alpha^\sharp
  \hook \omega \bigr)\,. 
\ee
We also use a shorthand $e^{a_1\dots a_p}=e^{a_1}\wedge \dots \wedge e^{a_p}$.

A {\em generalized twistor spinor} or {\em generalized conformal Killing spinor} ${\psi}$ is a spinor which for any vector field ${X}$ obeys the twistor equation with torsion: 
\be\label{ts}
{\nabla}^T_X {\psi}-\frac{1}{n}{X^\flat} D^T{\psi}=0\,.
\ee
Here the Dirac operator with torsion is defined as $D^T = e^a \nabla^T_{X_a} = D - \frac{3}{4} T\,,$
with $D$ being the Dirac operator of the Levi-Civita connection.
Similarly, we call a spinor $\psi$ obeying 
\be\label{ks}
{\nabla}^T_X {\psi}-\lambda{X}^\flat{\psi}=0
\ee
for some $\lambda\in \mathbb{C}$ a {\em generalized Killing spinor}.\footnote{%
It is easy to see, that a generalized twistor spinor $\psi$ which in addition obeys the Dirac equation with torsion, $D^T\psi=\frac{\lambda}{n}\psi$, is a generalized Killing spinor. }

\subsection{Twistor spinors and GCKY tensors}
Similar to the torsion-less case there is a connection between the existence of generalized twistor spinor and the existence of a tower of GCKY tensors. Namely, the following lemma holds: 
\begin{lemma}
Let ${\psi_1}$ and ${\psi_2}$ be two generalized twistor spinors. Then the $p$-form  ($p=1,\dots,n$)
\be 
\omega=(\psi_1, e^{a_1\dots a_p}\psi_2)e_{a_1\dots a_p}\,, 
\ee
where $(\cdot ,\cdot )$ stands for a spin-invariant symplectic product, is a GCKY tensor.
\end{lemma}
{\em Proof:} 
To prove this lemma we basically follow the calculation in App. A of \cite{Semmelmann:2002}. We calculate
\ba
\nabla^T_X\omega\!\!&=&\!\!\frac{1}{n}(X^\flat D^T\psi_1, e^{a_1\dots a_p}\psi_2)e_{a_1\dots a_p}+
\frac{1}{n}(\psi_1, e^{a_1\dots a_p}X^\flat D^T\psi_2)e_{a_1\dots a_p}\nonumber\\
&+&(\psi_1, \nabla^T_X[e^{a_1\dots a_p}]\psi_2)e_{a_1\dots a_p}+
(\psi_1, e^{a_1\dots a_p}\psi_2)\nabla^T_X e_{a_1\dots a_p}\nonumber\,.
\ea
To simplify our calculation, we can work in 
in a basis which is ``parallel at a point'', in which we have
\be
\nabla^T_X(e^{a_1\dots a_p})=\frac{p}{2}T(X,e^{[a_1},e_b)e^{|b|a_2\dots a_p]}\,.
\ee
Due to the antisymmetry of torsion $T$, we find that the last two terms cancel. Using further the property of the symplectic product $(\alpha u, v)=(-1)^{[q/2]}(u, \alpha v)$, valid for an arbitrary $q$-form $\alpha$, we arrive at 
\ba\label{b5}
\nabla^T_X\omega\!\!&=&\!\!\frac{1}{n}(D^T\psi_1, X^\flat e^{a_1\dots a_p}\psi_2)e_{a_1\dots a_p}+
\frac{1}{n}(\psi_1, e^{a_1\dots a_p}X^\flat D^T\psi_2)e_{a_1\dots a_p}\,.
\ea
Consider now
\ba
d^T\omega&=&e^b\wedge\nabla^T_{e_b}\omega\nonumber\\
&=& 
\frac{1}{n}(D^T\psi_1, e^{ba_1\dots a_p}\psi_2)e_{ba_1\dots a_p}+
\frac{(-1)^p}{n}(\psi_1, e^{ba_1\dots a_p}D^T\psi_2)e_{ba_1\dots a_p}\,.
\ea
So we get
\ba\label{b7}
X\hook d^T\omega&=&\frac{p+1}{n}(D^T\psi_1, X^\flat \wedge e^{a_1\dots a_p}\psi_2)e_{a_1\dots a_p}\nonumber\\
&&+(-1)^p
\frac{p+1}{n}(\psi_1, X^\flat \wedge e^{a_1\dots a_p}D^T\psi_2)e_{a_1\dots a_p}\,.
\ea
On the other hand, we have 
\ba
\delt\omega&=&-e_b\hook \nabla^T_{e_b}\omega\nonumber\\
&=& 
-\frac{p}{n}(D^T\psi_1, e_be^{ba_2\dots a_p}\psi_2)e_{a_2\dots a_p}-
\frac{p}{n}(\psi_1, e^{ba_2\dots a_p}e_bD^T\psi_2)e_{a_2\dots a_p}\nonumber\\
&=&
-\frac{p(n-p+1)}{n}(D^T\psi_1, e^{a_2\dots a_p}\psi_2)e_{a_2\dots a_p}\nonumber\\
&&+(-1)^p\frac{p(n-p+1)}{n}(\psi_1, e^{a_2\dots a_p}D^T\psi_2)e_{a_2\dots a_p}\,,\qquad
\ea
where we have used \eq{Cliff}. 
So we get 
\ba\label{b9}
X^\flat\wedge \delt\omega&=&
-\frac{n-p+1}{n}(D^T\psi_1, X\hook e^{a_1\dots a_p}\psi_2)e_{a_1\dots a_p}\nonumber\\
&&+(-1)^p\frac{n-p+1}{n}(\psi_1, X\hook e^{a_1\dots a_p}D^T\psi_2)e_{a_1\dots a_p}\,,
\ea
Putting \eq{b5}, \eq{b7}, and \eq{b9} together and using \eq{Cliff} again we finally obtain 
\ba
\nabla^T_X \omega-\frac{1}{p+1}{X}\hook {d}^T \omega+\frac{1}{n-p+1}
{X}^\flat \wedge {\delta}^T \omega=0\,.\qquad \qquad \Box \nonumber
\ea

\subsection{Special GCKY tensors}
Let as define a {\em special Killing--Yano tensor $\omega$ with torsion} to be a $p$-form which obeys
\be\label{SGKY}
\nabla^T_X \omega=\frac{1}{p+1}{X}\hook {d}^T \omega\,,\qquad
\nabla_X^T(d^T\omega)=c X^\flat \wedge \omega\,,
\ee
for any vector field $X$ and some constant $c$. It is a torsion generalization of a special Killing--Yano tensor introduced 
by Tachibana and Yu \cite{TachibanaYu:1970} and exploited by Semmelmann \cite{Semmelmann:2002}. 
Using \eq{SGKY} we immediately find that $\omega$ is an eigenform of the torsion Laplace-de Rham operator
\be
-(\dt\delt+\delt\dt)\omega=c(n-p)\omega\,.
\ee
Notice also that \eq{SGKY} implies $\dt\dt\omega=0$. 
Moreover, when $\omega$ is an odd-rank special Killing--Yano tensor with torsion, so is  ($k=0,1,\dots$) 
\be
\omega_{(k)}\equiv \omega\wedge (\dt\omega)^{\wedge k}\,.
\ee
Similarly, one can define a {\em special $d^T\!$-closed GCKY tensor} $\omega$ to be a $p$-form obeying 
\be\label{SGCKY}
\nabla^T_X \omega+\frac{1}{n-p+1}
{X}^\flat \wedge {\delta}^T \omega=0\,,\qquad 
\nabla_X^T(\delta^T\omega)=\tilde c X^\flat \hook \omega\,.
\ee
for any vector field $X$ and some constant $\tilde c$. Again, such $\omega$ is an eigenform of the torsion Laplace-de Rham operator,
$-(\dt\delt+\delt\dt)\omega=-\tilde cp\omega\,,$ and we have $\delt\delt\omega=0$.

Let us now consider a case when we have two generalized Killing spinors $\psi_1$ and $\psi_2$, 
\be
{\nabla}^T_X {\psi_1}-\lambda_1{X}^\flat{\psi_1}=0\,,\quad 
{\nabla}^T_X {\psi_2}-\lambda_2{X}^\flat{\psi_2}=0\,,
\ee
and construct a $p$-form $(p=1,\dots,n)$
\be 
\omega_p=(\psi_1, e^{a_1\dots a_p}\psi_2)e_{a_1\dots a_p}\,.
\ee
Then, by using 
\ba
X^\flat \wedge \omega_p\!\!&=&\!\!\frac{1}{p+1}(\psi_1,X\hook \wedge e^{a_1\dots a_{p+1}}\psi_2)e_{a_1\dots a_{p+1}}\,,\nonumber\\
X\hook \omega_p \!\!&=&\!\!p(\psi_1,X^\flat \wedge e^{a_1\dots a_{p-1}}\psi_2)e_{a_1\dots a_{p-1}}\,,\nonumber
\ea
and with the definition $k_{+}\equiv \bar \lambda_1+(-1)^p\lambda_2\,,\, k_{-}\equiv \bar \lambda_1-(-1)^p\lambda_2\,,$ we easily find that 
\be\label{b17}
\nabla_X^T\omega_p=\frac{k_+}{p+1}X\hook \omega_{p+1}+pk_- X^\flat \wedge \omega_{p-1}\,.
\ee
This means that $\omega_p$ is a GCKY $p$-form and moreover one has
\be
\dt\omega_p=k_+\omega_{p+1}\,,\qquad 
\delt\omega_p=-p(n-p+1)k_-\omega_{p-1}\,.
\ee
Taking a torsion derivative of these expressions while applying \eq{b17} we obtain
\ba
\nabla_X^{T}\!(\dt \omega_p)\!\!&=&\!\!\frac{k_+k_-}{p+2}X\hook \omega_{p+2}+k_+^2(p+1)X^\flat\wedge\omega_p\,,\\
\nabla_X^{T}\!(\delt \omega_p)\!\!&=&\!\!-(n-p+1)k_-^2X\hook \omega_{p}-p(p-1)(n-p+1)k_+k_-X^\flat\wedge\omega_p\,.
\ea
Obviously, when $k_-=0$ (which happens for example for $p$ odd and $\lambda_1=\lambda_2=\lambda\in  \mathbb{I}_m$), $\omega_p$ is a special Killing--Yano $p$-form with torsion, whereas when $k_+=0$, we have a special $d^T\!$-closed GCKY $p$-form. 
This allows us to formulate the following lemma, extending so the results of Cariglia \cite{Cariglia:2004}:
\begin{lemma}
Let ${\psi}$ be a generalized Killing spinor with purely imaginary $\lambda$, $\lambda\in  \mathbb{I}_m$. Then the above defined $\omega_p$ is a special Killing--Yano with torsion ($d^T\!$-closed GCKY) p-form for $p$ odd (even). Moreover, $\dt \omega_{2l+1}=-2\lambda\omega_{2l+2}$ and $\delt\omega=4\lambda(l+1)(n-2l-1)\omega_{2l+1}\,.$
\end{lemma}
A similar result (but with the words odd and even interchanged) is valid for $\lambda\in \mathbb{R}$.

\section{Bismut connection}
Let us consider a spacetime $(M,{g})$ 
admitting a PKY tensor ${h}$ with torsion.
With respect to the canonical frame $\{{e}^a\}$ 
the commutation relations are given by (\ref{comev1})--(\ref{comev4}) 
in even dimensions and (\ref{comod1})--(\ref{comod4}) in odd dimensions,
\begin{align}
\left[{e}_\mu,{e}_\nu \right] 
=& -\frac{x_\nu\sqrt{Q_\nu}}{x_\mu^2-x_\nu^2}\,{e}_\mu
    -\frac{x_\mu\sqrt{Q_\mu}}{x_\mu^2-x_\nu^2}\,{e}_\nu ~, \nonumber\\
\left[{e}_\mu,{e}_{\hat{\mu}}\right]
=& K_\mu\,{e}_\mu
  +L_\mu\,{e}_{\hat{\mu}}
  +\sum_{\rho\neq\mu}M_{\mu\rho}\,{e}_{\hat{\rho}}
  +\varepsilon\,J_\mu\,{e}_0 ~, \nonumber\\
\left[{e}_\mu,{e}_{\hat{\nu}}\right]
=&-\frac{x_\mu\sqrt{Q_\mu}}{x_\mu^2-x_\nu^2}\,{e}_{\hat{\nu}} ~, \nonumber\\
\left[{e}_{\hat{\mu}},{e}_{\hat{\nu}}\right]
=& 0 ~, \nonumber\\
\left[{e}_\mu,{e}_0\right]
=& -\frac{\sqrt{Q_\mu}}{x_\mu}\,{e}_0 ~, \nonumber\\
\left[{e}_{\hat{\mu}},{e}_0\right]
=& 0 ~,
\end{align}
where we have used Lemma 2.1.

In the case of even dimensions, we introduce an almost complex structure
\begin{equation}
{J}({e}_\mu)=-\chi_\mu {e}_{\hat{\mu}} ~,~~~ 
{J}({e}_{\hat{\mu}})=\frac{1}{\chi_\mu} {e}_{\mu} ~,
\end{equation}
where $\chi_\mu$ is an arbitrary function satisfying ${e}_\nu(\chi_\mu)={e}_{\hat{\nu}}(\chi_\mu)=0$ for $\nu\neq\mu$. 
Then the complex tangent space can be decomposed as $T^{C}{\cal M}={\cal D} \oplus {\cal \overline{D}}$
where ${\cal D}$ and ${\cal \overline{D}}$ are the eigenspaces 
corresponding to the eigenvalues $\sqrt{-1}$ and $-\sqrt{-1}$ respectively:
${\cal D}
= Span\big\{{e}_\mu+\sqrt{-1}\,\chi_\mu\,{e}_{\hat{\mu}}
   \,\big|\,\mu=1,\cdots,n\big\}$ and
${\cal \overline{D}}
= Span\big\{{e}_\mu-\sqrt{-1}\,\chi_\mu\,{e}_{\hat{\mu}}
   \,\big|\,\mu=1,\cdots,n\big\}$.
We find that the complex distribution ${\cal D}$
is integrable because $[{V},{W}]\in {\cal D}$ 
for any ${V}$, ${W}\in {\cal D}$,
which is equivalent to vanishing of the Nijenhuis tensor
\be
{N}({X},{Y}) 
\equiv [{J}({X}),~{J}({Y})]-[{X},~{Y}] -{J}([{X},~{J}({Y})])
        -{J}([{J}({X}),~{Y}])=0
\ee
for all ${X}, {Y} \in T{\cal M}$. 
Thus ${J}$ is a complex structure.
In particular, when we take $\chi_\mu=\epsilon_\mu$ with $\epsilon_\mu = \pm 1$, 
it is shown that the $2n$-dimensional spacetime $(M,\,{g})$
admits $2^n$ hermitian complex structures:
\begin{enumerate}
\item[(a)] ${J}_\epsilon = {J}\big|_{\chi_\mu=\epsilon_\mu}$ 
is a complex structure for each $\epsilon = (\epsilon_1,\cdots,\epsilon_n)$\,.
\item[(b)] $g$ is a hermitian metric, i.e., 
${g}({X}, {Y})={g}({J}_\epsilon {X}, {J}_\epsilon {Y})$\,.
\end{enumerate}

It is known \cite{Bismut:1989} 
that there exists a unique Hermitian connection $\nabla^B$ 
with a skew-symmetric torsion $B$, where a connection $\nabla^B$ is called Hermitian if
$\nabla^B {g}=0$, $\nabla^B {J}=0$. 
Hence we have a 2-form ${\omega}({X},{Y})={g}({X}, {J}({Y}))$ 
such that $\nabla^B {\omega}=0$.
This connection is called the {\it Bismut connection} 
and the corresponding manifold $(M,\,{g},\,{J},\,{\omega},\,{B})$
is called a {\it K\"ahler with torsion (KT) manifold}. 
By using such a 2-form ${\omega}$ the torsion can be written as
\begin{equation}
{B}({X},{Y},{Z})
=-d{\omega}({J}({X}),{J}({Y}),{J}({Z})) ~.
\end{equation} 
In the present case, the Bismut torsion associated with ${J}_\epsilon$ is explicitly given by
\begin{align}
 B_\epsilon = \sum_{\mu=1}^n\sum_{\nu\neq\mu}
  \Big(\frac{2\epsilon_\mu\epsilon_\nu x_\nu\sqrt{Q_\nu}}{x_\mu^2-x_\nu^2}-M_{\mu\nu}\Big)
  \,e^\mu\wedge e^{\hat{\mu}}\wedge e^{\hat{\nu}} ~. \label{Bismut}
\end{align}

The torsion ${T}$ associated with the PKY tensor ${h}$, 
cf.\ (\ref{formT}), 
is different from the Bismut torsion ${B}_\epsilon$.
If we choose as $\chi_\mu = x_\mu$ instead of $\chi_\mu=\epsilon_\mu$ for all $\mu$,
then the complex structure $\bar{{J}} = {J}\big|_{\chi_\mu=x_\mu}$ is naturally
related to the torsion ${T}$ as
\begin{align}
{T}({X},{Y},{Z})
=-d{h}(\bar{{J}}({X}),\bar{{J}}({Y}),\bar{{J}}({Z})) ~,
\end{align}
which gives a geometrical interpretation of (\ref{dh2}).

In the case of odd dimensions, 
we find a Cauchy-Riemann (CR) structure. Indeed, the complex distribution ${\cal D}_\epsilon$, 
where ${\cal D}_\epsilon=Span\{{e}_\mu+\epsilon_\mu\sqrt{-1}\,{e}_{\hat{\mu}}
\,|\,\mu=1,\cdots,n\}\subset T^{C}{\cal M}$, 
is integrable because $[{Z},{W}]\in {\cal D}_\epsilon$ 
for any ${Z}, {W}\in {\cal D}_\epsilon$.

\section{Covariant derivatives}
In this appendix, we gather covariant derivatives $\nabla^T_{e_a}{e}_b$. These were calculated using Lemma~3.1 in even dimensions 
and Lemma~4.1 in odd dimensions. Integrability conditions of the PKY equation (\ref{eq311}) and (\ref{4.13}) have also been employed.
As a results, $\nabla^T_{e_a}{e}_b$ are determined in terms of the PKY eigenvalues $x_\mu$, unknown functions $Q_\mu$ and $Q_0$, and derivatives of the associated 1-form $\kappa_a{}^b$ defined by (\ref{kappadef}). 
We have the following results:
\begin{enumerate}
\item In even dimensions
\begin{align}
\nabla^T_{e_\mu}{e}_\mu
=& \sum_{\rho\neq\mu}\frac{x_\rho\sqrt{Q_\rho}}{x_\mu^2-x_\rho^2}\,{e}_\rho
   -\frac{\kappa_\mu{}^\mu}{\sqrt{Q_\mu}}\,{e}_\mu ~, \nonumber\\
\nabla^T_{e_\mu}{e}_\nu
=& -\frac{x_\nu\sqrt{Q_\nu}}{x_\mu^2-x_\nu^2}\,{e}_\mu ~~(\mu\neq\nu) ~, \nonumber\\
\nabla^T_{e_\mu}{e}_{\hat{\mu}}
=& \sum_{\rho\neq\mu}\frac{x_\mu\sqrt{Q_\rho}}{x_\mu^2-x_\rho^2}\,{e}_{\hat{\rho}}
  +\frac{\kappa_\mu{}^\mu}{\sqrt{Q_\mu}}\,{e}_{\hat{\mu}} ~, \nonumber\\
\nabla^T_{e_\mu}{e}_{\hat{\nu}}
=& -\frac{x_\mu\sqrt{Q_\nu}}{x_\mu^2-x_\nu^2}\,{e}_{\hat{\mu}} ~~(\mu\neq\nu) ~, \nonumber\\
\nabla^T_{e_{\hat{\mu}}}{e}_\mu
=& \frac{1}{\sqrt{Q_\mu}}\Bigg(\sum_{\rho\neq\mu}
  \frac{x_\mu Q_\rho}{x_\mu^2-x_\rho^2}-\kappa_{\hat{\mu}}{}^\mu\Bigg)\,{e}_{\hat{\mu}} 
  -\sum_{\rho\neq\mu}\frac{x_\mu\sqrt{Q_\rho}}{x_\mu^2-x_\rho^2}\,{e}_{\hat{\rho}} ~, \nonumber\\
\nabla^T_{e_{\hat{\mu}}}{e}_\nu
=& -\frac{x_\nu\sqrt{Q_\nu}}{x_\mu^2-x_\nu^2}\,{e}_{\hat{\mu}}
  +\Bigg(\frac{x_\nu\sqrt{Q_\mu}}{x_\mu^2-x_\nu^2}
  -\frac{\kappa_{\hat{\mu}}{}^\nu}{\sqrt{Q_\nu}}\Bigg)\,{e}_{\hat{\nu}} ~~(\mu\neq\nu) ~, \nonumber\\
\nabla^T_{e_{\hat{\mu}}}{e}_{\hat{\mu}}
=& -\frac{1}{\sqrt{Q_\mu}}\Bigg(\sum_{\rho\neq\mu}
   \frac{x_\mu Q_\rho}{x_\mu^2-x_\rho^2}-\kappa_{\hat{\mu}}{}^\mu\Bigg)\,{e}_\mu
 +\sum_{\rho\neq\mu}\frac{x_\rho\sqrt{Q_\rho}}{x_\mu^2-x_\rho^2}\,{e}_\rho ~,\nonumber\\
\nabla^T_{e_{\hat{\mu}}}{e}_{\hat{\nu}}
=& \frac{x_\mu\sqrt{Q_\nu}}{x_\mu^2-x_\nu^2}\,{e}_\mu
  -\Bigg(\frac{x_\nu\sqrt{Q_\mu}}{x_\mu^2-x_\nu^2}
  -\frac{\kappa_{\hat{\mu}}{}^\nu}{\sqrt{Q_\nu}}\Bigg)\,{e}_\nu ~~(\mu\neq\nu) ~.
\end{align}
 \item In odd dimensions
\begin{align}
\nabla^T_{e_\mu}{e}_\mu
=& \sum_{\rho\neq\mu}\frac{x_\rho\sqrt{Q_\rho}}{x_\mu^2-x_\rho^2}\,{e}_\rho
   -\frac{\kappa_\mu{}^\mu}{\sqrt{Q_\mu}}\,{e}_{\hat{\mu}} ~, \nonumber\\
\nabla^T_{e_\mu}{e}_\nu
=& -\frac{x_\nu\sqrt{Q_\nu}}{x_\mu^2-x_\nu^2}\,{e}_\mu ~~(\mu\neq\nu) ~, \nonumber\\
\nabla^T_{e_\mu}{e}_{\hat{\mu}}
=& \sum_{\rho\neq\mu}\frac{x_\mu\sqrt{Q_\rho}}{x_\mu^2-x_\rho^2}\,{e}_{\hat{\rho}}
  +\frac{\kappa_\mu{}^\mu}{\sqrt{Q_\mu}}\,{e}_{\hat{\mu}}
  +\frac{\sqrt{Q_0}}{x_\mu}\,{e}_0 ~, \nonumber\\
\nabla^T_{e_\mu}{e}_{\hat{\nu}}
=& -\frac{x_\mu\sqrt{Q_\nu}}{x_\mu^2-x_\nu^2}\,{e}_{\hat{\mu}} ~~(\mu\neq\nu) ~, \nonumber\\
\nabla^T_{e_{\hat{\mu}}}{e}_\mu
=& \frac{1}{\sqrt{Q_\mu}}\Bigg(\sum_{\rho\neq\mu}
  \frac{x_\mu Q_\rho}{x_\mu^2-x_\rho^2}+\frac{Q_0}{x_\mu}-\kappa_{\hat{\mu}}{}^\mu\Bigg)\,{e}_{\hat{\mu}}
  -\sum_{\rho\neq\mu}\frac{x_\mu\sqrt{Q_\rho}}{x_\mu^2-x_\rho^2}\,{e}_{\hat{\rho}}
  -\frac{\sqrt{Q_0}}{x_\mu}\,{e}_0 ~, \nonumber\\
\nabla^T_{e_{\hat{\mu}}}{e}_\nu
=& -\frac{x_\nu\sqrt{Q_\nu}}{x_\mu^2-x_\nu^2}\,{e}_{\hat{\mu}}
  +\Bigg(\frac{x_\nu\sqrt{Q_\mu}}{x_\mu^2-x_\nu^2}
  -\frac{\kappa_{\hat{\mu}}{}^\nu}{\sqrt{Q_\nu}}\Bigg)\,{e}_{\hat{\nu}} ~~(\mu\neq\nu) ~, \nonumber\\
\nabla^T_{e_{\hat{\mu}}}{e}_{\hat{\mu}}
=& -\frac{1}{\sqrt{Q_\mu}}\Bigg(\sum_{\rho\neq\mu}
  \frac{x_\mu Q_\rho}{x_\mu^2-x_\rho^2}+\frac{Q_0}{x_\mu}-\kappa_{\hat{\mu}}{}^\mu\Bigg)\,{e}_\mu
  +\sum_{\rho\neq\mu}\frac{x_\rho\sqrt{Q_\rho}}{x_\mu^2-x_\rho^2}\,{e}_\rho ~, \nonumber\\
\nabla^T_{e_{\hat{\mu}}}{e}_{\hat{\nu}}
=& \frac{x_\mu\sqrt{Q_\nu}}{x_\mu^2-x_\nu^2}\,{e}_\mu
  -\Bigg(\frac{x_\nu\sqrt{Q_\mu}}{x_\mu^2-x_\nu^2}
  -\frac{\kappa_{\hat{\mu}}{}^\nu}{\sqrt{Q_\nu}}\Bigg)\,{e}_\nu ~~(\mu\neq\nu) ~, \nonumber\\
\nabla^T_{e_\mu}{e}_0
=& -\frac{\sqrt{Q_0}}{x_\mu}\,{e}_{\hat{\mu}} ~, \nonumber\\
\nabla^T_{e_{\hat{\mu}}}{e}_0
=& \frac{\sqrt{Q_0}}{x_\mu}\,{e}_\mu ~, \nonumber\\
\nabla^T_{e_0}{e}_\mu
=& -\Bigg(\frac{\sqrt{Q_0}}{x_\mu}+\frac{\kappa_0{}^\mu}{\sqrt{Q_\mu}}\Bigg)
  \,{e}_{\hat{\mu}}
  +\frac{\sqrt{Q_\mu}}{x_\mu}\,{e}_0 ~, \nonumber\\
\nabla^T_{e_0}{e}_{\hat{\mu}}
=& \Bigg(\frac{\sqrt{Q_0}}{x_\mu}+\frac{\kappa_0{}^\mu}{\sqrt{Q_\mu}}\Bigg) \,{e}_\mu ~, \nonumber\\
\nabla^T_{e_0}{e}_0
=& -\sum_{\mu=1}\frac{\sqrt{Q_\mu}}{x_\mu}\,{e}_\mu ~.
\end{align}
\end{enumerate}


\end{document}